%% file: JournalMain.tex
\documentclass[manuscript, screen]{jair}

\usepackage{fancybox}
\usepackage{todonotes}

\RequirePackage[
  datamodel=acmdatamodel,
  style=acmauthoryear,
  backend=biber,
  giveninits=true
  ]{biblatex}

\renewcommand\footnotetextcopyrightpermission[1]{}

\JAIRTrack{}   
\JAIRAE{}      

\begin{document}

\pagestyle{plain}

\title[Symbolic Functional Decomposition]{Symbolic Functional Decomposition: A Reconfiguration Approach}

\author{Mateus de Oliveira Oliveira}
\authornote{Both authors contributed equally to this research.}
\orcid{0000-0001-7798-7446}
\email{oliveira@dsv.su.se}

\affiliation{%
  \institution{Stockholm University}
  \country{Sweden}
}
\affiliation{%
  \institution{University of Bergen}
  \country{Norway}
}

\author{Wim Van den Broeck}
\authornotemark[1]
\orcid{0000-0002-5305-7348}
\affiliation{%
  \institution{University of Bergen}
  \country{Norway}}
\email{wim.broeck@uib.no}

\renewcommand{\shortauthors}{Van den Broeck, de Oliveira Oliveira}
\input{JournalPreamble}
\input{00-Abstract}




\maketitle
\thispagestyle{plain}

\input{01-Introduction}
\input{02-Preliminaries}
\input{03-AutomataVsFunctionClasses}
\input{04-FunctionalReconfiguration}
\input{ProofOfMainTheorem}

\input{05-Applications}
\input{06-ConclusionAndRelatedWork}

\begin{acks}
    We acknowledge support from the Research Council of Norway, grant numbers 288761 and 326537. 
\end{acks}
\printbibliography

\appendix

\section{Omitted proofs}\label{appedixOmitted}

We present the proof of Corollary \ref{corollary:RegularityRelations} restated below as Corollary \ref{corollary:RegularityRelationsProof}.
\begin{corollary}
\label{corollary:RegularityRelationsProof}
Let  $\widthone,\widthtwo,\widththree\in \N$, and 
$\width$ be equal to $\max\{\widthone,\widthtwo\}$ (or $\max\{\widthone,\widthtwo,\widththree\}$ where appropriate).
\begin{enumerate}
\item $\relation_{=}(\widthone,\widthtwo)$ is $2^{\mathcal{O}(\width \cdot \log\width)}$-regular. 
\item $\relation_{\cap}(\widthone,\widthtwo,\widththree)$ is $2^{\mathcal{O}(\width^2\log\width)}$-regular.  
\item $\relation_{\cup}(\widthone,\widthtwo,\widththree)$ is $2^{\mathcal{O}(\width^2\log\width)}$-regular. 
\item $\relation_{\neg}(\widthone,\widthtwo)$ is  $2^{\mathcal{O}(\width \cdot \log\width)}$-regular
\end{enumerate}
\end{corollary}

\begin{proof}
We give a compact description of an automata realizing each relation of the Corollary. As in the statement of the Corollary we will take $\width$ to be equal to $\max\{\widthone,\widthtwo\}$ or $\max\{\widthone,\widthtwo,\widththree\}$ where appropriate. 

We denote by $\relation_\mathrm{id}(\w)$ the relation consisting of pairs of identical OBDDs, $\{(\D,\D)\;|\; \D\in 
\detCalB(\width)^\circledast\}$. This relation is $2^{\w}$ by a straightforward adaptation of the proof of Proposition \ref{proposition:RegularityOBDDs}.\\

\noindent$\boldsymbol{\relation_{=}(\widthone,\widthtwo)}$:
Let $\languageL = \Lang(\relation^{\mathrm{det}}_\mathrm{can}(\widthone, \widthtwo)) \otimes \Lang(\relation^{\mathrm{det}}_\mathrm{can}(\widthone,\widthtwo))$. Since $\relation^{\mathrm{det}}_\mathrm{can}(\widthone,\widthtwo)$ is $2^{\mathcal{O}(\w \cdot \log\w)}$-regular, $\languageL$ is $2^{\mathcal{O}(\w \cdot \log\w)}$-regular by Proposition \ref{proposition:TensorProduct}. Let $(\relation,\selectorTuple)$ be a $(2,4)$-selector where $\relation=\relation_\mathrm{id}(\w)$ and $\selectorTuple=\{2,4\}$. The $(\relation,\selectorTuple)$-selection $\selection{\languageL}{\relation}{\selectorTuple}$ is $2^{\mathcal{O}(\w \cdot \log\w)}$-regular by the selection lemma (Proposition \ref{proposition:SelectionLemma}). Now consider projection $\pi:\widehat{\calB}(\w)^{\times 4} \rightarrow \widehat{\calB}(\w)^{\times 2}$, where $\pi(\B_1,\B_2,\B_3,\B_4)=(\B_1,\B_3)$. We note that projecting the first and third coordinates of $\selection{\languageL}{\relation}{\selectorTuple}$ gives $\pi(\selection{\languageL}{\relation}{\selectorTuple})=\Lang(\relation_{=}(\widthone,\widthtwo))$ which remains $2^{\mathcal{O}(\w \cdot \log\w)}$-regular.\\

\noindent$\boldsymbol{R_{\cap}(\widthone,\widthtwo,\widththree)}$: Letting $\gamma:\dbset{\w}\times\dbset{\w} \rightarrow \dbset{\w^2}$ be a bijective function such that $\gamma(0,0)= 0$, we define first function $f_\gamma: \detCalB(\w) \times \detCalB(\w) \rightarrow \detCalB(\w^2)$,
$$
f_\gamma(\B_i,\B_i') = \{(\gamma(q,q'), a, \gamma(p,p')) \;|\; (q,a,p) \in \B_i, (q',a,p') \in \B_i'\}
$$
This function bijectively maps pairs of layers in $\detCalB(\w)$ to a single layer in $\detCalB(\w^2)$.

Next let $\gamma': \dbset{\w} \times \dbset{\w} \rightarrow \dbset{\w^2}$ be a function such that for $x,y \in \dbset{\w}$, $\gamma'(x,y)=0$ if and only if $x=0$ or $y=0$. We now define $f_{\gamma'}: \detCalB(\w) \times \detCalB(\w) \rightarrow \detCalB(\w^2)$; a slight modification of $f_\gamma$.
$$
f_{\gamma'}(\B_i,\B_i') = \{(\gamma(q,q'), a, \gamma'(p,p')) \;|\; (q,a,p) \in \B_i, (q',a,p') \in \B_i'\}
$$

This function also maps pairs of layers in $\detCalB(\w)$ to a single layer in $\detCalB(\w^2)$ but differs from $f_\gamma$ in that for $x,y\in\dbset{\w}$ where $x=0$ or $y=0$, $f_{\gamma'}$ collapses states of the form $\gamma'(x,y)$, to a single state, $0$ in $\dbset{\w^2}$. In other words, $f_{\gamma'}$ ensures that if $f_{\gamma'}(\B_i,\B_i')$ is the final layer of an ODD $f_{\gamma'}(\D,\D')$ it accepts only those strings that are accepted by both $\D$ and $\D'$. 

We note now that given two ODD's $\D = \B_1\B_2\ldots\B_l$ and $\D'= \B'_1\B'_2\ldots\B'_l$ we can obtain ODD $\D''= f_\gamma(\B_1,\B_1') f_\gamma(\B_2,\B_2') \ldots f_\gamma(\B_{l-1},\B'_{l-1}) f_{\gamma'}(\B_l,\B_l')$. The ODD $\D''$ is an ODD representing the intersection of the languages of $\D$ and $\D'$, $\Lang(\D'') = \Lang(\D) \cap \Lang(\D')$.

Using functions $f_{\gamma}$and $f_{\gamma'}$ we can now define the finite automaton $\automaton$ over alphabet $\widehat{\calB}(\widthone) \otimes \widehat{\calB}(\widthtwo) \otimes \widehat{\calB}(\widthone \cdot \widthtwo) \otimes \widehat{\calB}(\widththree)$ as follows. The automaton will have state set $Q(\automaton)=Q_a \cup Q_b$ where sets $Q_a$ and $Q_b$ are defined as follows:
\begin{itemize}
    \item $Q_a = \{q_{[S_1,S_2,S_3]} \;|\; S_1 \subseteq \dbset{\widthone},S_2 \subseteq \dbset{\widthtwo},S_3 \subseteq \dbset{\widththree}\}$
    \item $Q_b = \{q'_{[S_1,S_2,S_3]} \;|\; S_1 \subseteq \dbset{\widthone},S_2 \subseteq \dbset{\widthtwo},S_3 \subseteq \dbset{\widththree}\}$
\end{itemize}

The state $q_{\{0\},\{0\},\{0\}}$ is considered the initial state, $I(\automaton) = q_{\{0\},\{0\},\{0\}}$, while set of states $Q_b$ are considered final states, $F(\automaton)=Q_b$. The automata has transition set $T(\automaton)= T_a \cup T_b$ where sets $T_a$ and $T_b$ are defined as follows: 
\begin{itemize}
    \item $T_a =\{(q_{[S_1',S_2',S_3']}, (\B_1, \B_2, f_\gamma(\B_1,\B_2),\B_3), q_{[S_1,S_2,S_3]}) \;|\; \Dom(\B_i)=S_i', \im(\B_i)=S_i\}$
    \item $T_b =\{(q_{[S_1',S_2',S_3']}, (\B_1, \B_2, f'_\gamma(\B_1,\B_2),\B_3), q'_{[S_1,S_2,S_3]}) \;|\; \Dom(\B_i)=S_i', \im(\B_i)=S_i\}$
\end{itemize}

One can readily verify that, $\Lang(\automaton)$ is the language of tensor products of 4 ODDs $\D_1 \otimes\D_2 \otimes \D_3 \otimes \D_4$, where $\D_1 \in \widehat{\calB}(\widthone),\D_2 \in \widehat{\calB}(\widthtwo),\D_3 \in \widehat{\calB}(\widthone \cdot \widthtwo), \D_4\in \widehat{\calB}(\widththree)$ and $\Lang(\D_3) = \Lang(\D_1) \cap \Lang(\D_2)$. The language $\Lang(\automaton)$ is
$2^{\mathcal{O}(\w)}$-regular. Let $(\relation,\selectorTuple)$ be a $(\widehat{\calB}(\w^2),2,4)$-selector where $\relation=\relation_\mathrm{=}(\w^2)$ and $\selectorTuple=\{3,4\}$. The $(\relation,\selectorTuple)$-selection $\selection{\languageL}{\relation}{\selectorTuple}$ is $2^{\mathcal{O}(\w^2 \cdot \log\w)}$-regular by the selection lemma (Proposition \ref{proposition:SelectionLemma}). Now consider projection $\pi:\widehat{\calB}(\w^2)^{\times 4} \rightarrow \widehat{\calB}(\w^2)^{\times 3}$, where $\pi(\B_1,\B_2,\B_3,\B_4)=(\B_1,\B_2,\B_4)$. We note that projecting away the third coordinate of $\selection{\languageL}{\relation}{\selectorTuple}$ gives $\pi(\selection{\languageL}{\relation}{\selectorTuple})=\Lang(\relation_{\cap}(\widthone,\widthtwo,\widththree))$ which remains $2^{\mathcal{O}(\w^2 \cdot \log\w)}$-regular.\\

\noindent$\boldsymbol{\relation_{\cup}(\widthone,\widthtwo,\widththree)}$:
Similar to the previous case, we let  and define $\gamma:\dbset{\w}\times\dbset{\w} \rightarrow \dbset{\w^2}$ be a bijective function such that $\gamma(0,0)= 0$ and define function $f_\cup:\detCalB(\w) \times \detCalB(\w) \rightarrow \detCalB(\w^2)$ where,
$$ f_\cup(\B,\B') = \{(\gamma(q,q'), a, \gamma(p,p')) \;|\; (q,a,p) \in \B, (q',a,p') \in \B'\}$$

We note now that given two ODD's $\D = \B_1\B_2\ldots\B_l$ and $\D'= \B'_1\B'_2\ldots\B'_l$ we can obtain ODD $\D''= f_\cup(\B_1,\B_1') f_\cup(\B_2,\B_2') \ldots f_\cup(\B_l,\B_l')$. The ODD $\D''$ is an ODD representing the union of the languages of $\D$ and $\D'$, $\Lang(\D'') = \Lang(\D) \cup \Lang(\D')$.

Using function $f_\cup$ we can now define the following finite automaton $\automaton$ over alphabet $\widehat{\calB}(\widthone) \otimes \widehat{\calB}(\widthtwo) \otimes \widehat{\calB}(\widthone \cdot \widthtwo) \otimes \widehat{\calB}(\widththree)$. The state set $Q(\automaton)$ is defined as follows:
$$Q(\automaton)=\{q_{[S_1,S_2,S_3]} \;|\; S_1 \subseteq \dbset{\widthone},S_2 \subseteq \dbset{\widthtwo},S_3 \subseteq \dbset{\widththree}\}$$
The state $q_{\{0\},\{0\},\{0\}}$ is considered the initial state, $I(\automaton) = q_{\{0\},\{0\},\{0\}}$, while all other states are considered final states, $F(\automaton)=Q(\automaton)\setminus I(\automaton)$. The transition set $T(\automaton)$ is defined as follows:
$$T(\automaton) = \{(q_{[S_1',S_2',S_3']}, (\B_1, \B_2, f_\cup(\B_1,\B_2),\B_3), q_{[S_1,S_2,S_3]}) \;|\; \Dom(\B_i)=S_i', \im(\B_i)=S_i\} $$

The language $\Lang(\automaton)$ is $2^{\mathcal{O}(\w)}$-regular. Let $(\relation,\selectorTuple)$ be a $(\widehat{\calB}(\w^2),2,4)$-selector where $\relation=\relation_\mathrm{=}(\w^2)$ and $\selectorTuple=\{3,4\}$. The $(\relation,\selectorTuple)$-selection $\selection{\languageL}{\relation}{\selectorTuple}$ is $2^{\mathcal{O}(\w^2 \cdot \log\w)}$-regular by the selection lemma (Proposition \ref{proposition:SelectionLemma}). Now consider projection $\pi:\widehat{\calB}(\w^2)^{\times 4} \rightarrow \widehat{\calB}(\w^2)^{\times 3}$, where $\pi(\B_1,\B_2,\B_3,\B_4)=(\B_1,\B_2,\B_4)$. We note that projecting away the third coordinate of $\selection{\languageL}{\relation}{\selectorTuple}$ gives $\pi(\selection{\languageL}{\relation}{\selectorTuple})=\Lang(\relation_{\cup}(\widthone,\widthtwo,\widththree))$ which remains $2^{\mathcal{O}(\w^2 \cdot \log\w)}$-regular.\\

\noindent{$\boldsymbol{\relation_{\neg}(\widthone,\widthtwo)}$:} 
We define first the function $f_\neg:\detCalB(\w) \rightarrow \detCalB(\w)$ where,
$$
\begin{array}{l}
     f_\neg(\B) = \{(q, a, 1) \;|\; (q,a,p) \in \B, p = 0\} \cup 
     \{(q, a, 0) \;|\; (q,a,p) \in \B, p \in \dbset{\w}\setminus\{0\}\}
\end{array}
$$

This function redirects transitions mapping to state $0$ such that they map to state $1$ and redirects transitions not mapping to $0$ such that the map to $0$. Using $f_\neg(\B)$ we can define the following finite automaton $\automaton$ over alphabet $\widehat{\calB}(\w)^{\times 3}$. The automaton has state set $Q(\automaton)= Q_a \cup Q_b$ where sets $Q_a$ and $Q_b$ are defined as follows:
\begin{itemize}
    \item $Q_a =\{q_{[S_1,S_2]} \;|\; S_1 \subseteq \dbset{\widthone},S_2 \subseteq \dbset{\widthtwo}\}$
    \item $Q_b =\{q'_{[S_1,S_2]} \;|\; S_1 \subseteq \dbset{\widthone},S_2 \subseteq \dbset{\widthtwo}\}$
\end{itemize}

The state $q_{\{0\},\{0\}}$ is considered the initial state, $I(\automaton) = q_{\{0\},\{0\}}$, while set of states $Q_b$ are considered final states, $F(\automaton)=Q_b$. The automata has transition set $T(\automaton)= T_a \cup T_b$ where sets $T_a$ and $T_b$ are defined as follows:
\begin{itemize}
    \item $T_a = \{(q_{[S_1',S_2']}, (\B_1, \B_1,\B_2), q_{[S_1,S_2]}) \;|\; \Dom(\B_i)=S_i', \im(\B_i)=S_i\}$
    \item $T_b =\{(q_{[S_1',S_2']}, (\B_1,f_\neg(\B_1),\B_2), q'_{[S_1,S_2]}) \;|\; \Dom(\B_i)=S_i', \im(\B_i)=S_i\}$
\end{itemize}

The language $\Lang(\automaton)$ is $2^{\mathcal{O}(\w)}$-regular. Let $(\relation,\selectorTuple)$ be a $(\widehat{\calB}(\w),2,3)$-selector where $\relation=\relation_\mathrm{=}(\w)$ and $\selectorTuple=\{2,3\}$. The $(\relation,\selectorTuple)$-selection $\selection{\languageL}{\relation}{\selectorTuple}$ is $2^{\mathcal{O}(\w \cdot \log\w)}$-regular by the selection lemma (Proposition \ref{proposition:SelectionLemma}). Now consider projection $\pi:\widehat{\calB}(\w)^{\times 3} \rightarrow \widehat{\calB}(\w)^{\times 2}$, where $\pi(\B_1,\B_2,\B_3)=(\B_1,\B_3)$. We note that projecting away the third coordinate of $\selection{\languageL}{\relation}{\selectorTuple}$ gives $\pi(\selection{\languageL}{\relation}{\selectorTuple})=\Lang(\relation_{\neg}(\widthone,\widthtwo))$ which remains $2^{\mathcal{O}(\w \cdot \log\w)}$-regular.

\end{proof}

\end{document}

%% file: JournalPreamble.tex
\newcommand{\eqdef}{=}
\newcommand{\N}{\mathbb{N}}
\newcommand{\ngates}{m}
\newcommand{\stringx}{x} 
\newcommand{\stringy}{y}
\newcommand{\s}{s} 
\newcommand{\dbset}[1]{\llbracket #1 \rrbracket }
\newcommand{\Nplus}{\mathbb{N}_+}

\newcommand{\lengthString}{n}

\newcommand{\paddingSymbol}{\square}
\newcommand{\symbola}{a}

\newcommand{\allfunctions}{\mathcal{F}}

\newcommand{\D}{D} 
\newcommand{\B}{B} 
\newcommand{\calB}{\mathcal{B}} 
\newcommand{\detCalB}{\widehat{\mathcal{B}}}
\newcommand{\w}{w} 
\newcommand{\width}{w} 
\newcommand{\widthone}{w_1}
\newcommand{\widthtwo}{w_2}
\newcommand{\widththree}{w_3}
\newcommand{\Dom}{\mathrm{Dom}}
\newcommand{\im}{\mathrm{Im}}
\newcommand{\Det}{\mathrm{Det}}
\newcommand{\can}{\mathcal{C}}
\newcommand{\ODDwidth}{\mathrm{w}}

\newcommand{\canonicalOBDD}{\mathsf{Can}}
\newcommand{\classfunctions}{\mathcal{C}}

\newcommand{\gate}{g} 
\newcommand{\circuit}{C}
\newcommand{\selectorlanguage}[2]{\mathfrak{S}[#1,#2]}
\newcommand{\selection}[3]{\mathrm{sel}(#1,#2,#3)}
\newcommand{\Lang}{\mathcal{L}}
\newcommand{\Sw}{\mathrm{Sw}}
\newcommand{\arityvalue}{r}
\newcommand{\emptystring}{\lambda}
\newcommand{\p}{p}
\newcommand{\alphabet}{\Sigma}
\newcommand{\decomposerFunction}{F}
\newcommand{\decomposerCircuit}{C}

\newcommand{\gfunction}{g}
\newcommand{\ffunction}{f}
\newcommand{\ncomponents}{k}
\newcommand{\inputlength}{n}

\newcommand{\gatefunction}{F}
\newcommand{\transformation}{\mathrm{Tr}}
\newcommand{\widthtransformation}{\mathbf{w}}
\newcommand{\transformationwidth}{\mathsf{trw}}

\newcommand{\classfunctionsG}{\mathcal{G}}
\newcommand{\classfunctionsF}{\mathcal{F}}

\newcommand{\selectorTuple}{S}

\newcommand{\functionclassautomaton}{\mathcal{A}}
\newcommand{\automaton}{\mathcal{A}}
\newcommand{\states}{Q}
\newcommand{\initialStates}{I}
\newcommand{\finalStates}{F}
\newcommand{\transitions}{\Delta}
\newcommand{\stateq}{q}
\newcommand{\languageL}{L}
\newcommand{\regularityalpha}{\alpha}
\newcommand{\regularitybeta}{\beta}

\newcommand{\mapAlphabet}{\mathfrak{m}}
\newcommand{\alphabetGamma}{\Gamma}
\newcommand{\symbolSigma}{\sigma}
\newcommand{\relation}{R}

\newcommand{\parproblem}[4]{\begin{center}\shadowbox{\begin{minipage}{.95\linewidth}
				{{{\bf Problem name:} {\sc #1}{\index{\sc #1}}} 
				\\
				{\bf Given:} #2\\
				{\bf Parameters:} #3\\ 
				{\bf Question:} #4}
			\end{minipage}}
	\end{center}}

\newcommand{\circuitsize}{m}
\newcommand{\pnFunctionDecomposition}{{\sc FunctionalDecomposition~}}
\newcommand{\pnGenJunta}{{\sc GeneralizedJunta~}}
\newcommand{\pnRecFunctionDecomposition}{{\sc FunctionalReconfiguration~}}

\newcommand{\circuitDepth}{d}
\newcommand{\complexityfunction}{h}
\newcommand{\solutionWidth}{p}
\newcommand{\variable}{x}

\newcommand{\depth}{d}
\newcommand{\consistencyLanguage}{\mathit{Con}}

\newtheorem{observation}[theorem]{Observation}
\newtheorem{question}[theorem]{Question}

\newcommand{\solutionLanguage}{\mathit{Sol}}
\newcommand{\hypercubeautomaton}{\mathcal{H}}

%% file: 00-Abstract.tex
\begin{abstract}
Functional decomposition is the process of breaking down a function $f$ into a composition $f=g(f_1,\dots,f_k)$ of simpler functions $f_1,\dots,f_k$ belonging to some class $\mathcal{F}$. This fundamental notion can be used to model applications arising in a wide variety of contexts, ranging from machine learning to formal language theory. 
In this work, we study functional decomposition by leveraging on the notion of functional reconfiguration. In this setting, constraints are imposed not only on the factor functions $f_1,\dots,f_k$ but also on the intermediate functions arising during the composition process. 

We introduce a symbolic framework to address functional reconfiguration and decomposition problems. In our framework, functions arising during the reconfiguration process are represented symbolically, using ordered binary decision diagrams (OBDDs). The function $g$  used to specify the reconfiguration process is represented by a Boolean circuit $C$. Finally, the function class $\mathcal{F}$ is represented by a second-order finite automaton $\mathcal{A}$. Our main result states that functional reconfiguration, and hence functional decomposition, can be solved in fixed-parameter linear time when parameterized by the width of the input OBDD, by structural parameters associated with the reconfiguration circuit $C$, and by the size of the second-order finite automaton $\mathcal{A}$.
\end{abstract}

%% file: 01-Introduction.tex
\section{Introduction}
\label{section:Introduction}

Cognition and learning are processes that often occur in a hierarchical manner, where complex concepts are broken down into simpler, more manageable ones \cite{DBLP:conf/ijcai/GoelC89,correa2023humans,DBLP:conf/nips/AtzmonKSC20,DBLP:conf/ijcai/WangYWD23}.
In contexts where concepts can be represented by functions, this hierarchical approach to processing and understanding information can often be formalized using an appropriate notion of functional decomposition \cite{zupan1997machine,bohanec2004function,vykhovanets2006algebraic,deniziak2021decolib}. 

In this work, we view functional decomposition as the process of breaking down a Boolean function $\ffunction$ into a composition $g(\ffunction_1,\dots,\ffunction_{\ncomponents})$, where $\ffunction_1,\dots,\ffunction_{\ncomponents}$ are functions from a pre-defined class $\classfunctionsF$. The function $g$ specifies how the component functions $\ffunction_1,\dots,\ffunction_{\ncomponents}$ combine to form the target function $\ffunction$, while $\classfunctionsF$ is a class of functions that are considered to be simpler than the target function $\ffunction$. For instance, $\mathcal{F}$ may be a class of functions whose preimage is closed under some algebraic operation, or a class of functions of bounded complexity with respect to some suitable complexity measure. 

Functional decomposition is a fundamental concept within the realm of artificial intelligence because it promotes modularity, interpretability and explainability. First, functional decomposition enables the reuse of well-defined components across distinct AI systems \cite{DBLP:conf/ijcai/SunNM19a,DBLP:conf/ijcai/IzzaJRB20,DBLP:journals/jair/Srivastava23}. In this context one could hope to implement a system with a given functionality $f$ by combining pre-existing systems with functionality $f_1,f_2,\dots,f_k$. Second, decomposing a complex function into sub-functions makes the model more interpretable. Each sub-function can be understood as performing a specific task, or transformation, which provides insights on how the model arrives at its final computations. Finally, it helps in explaining the decision-making process of computational models by breaking down the decision into simpler, easier to understand steps \cite{molnar2020quantifying,zupan1999function,hashemi2018evolutionary,hiabu2023unifying,laberge2024tackling}.

As a closely related concept, in this work we also consider the notion of {\em functional reconfiguration}. In this setting, not only 
the factor functions $f_1,\dots,f_k$ are required to satisfy some property, but also the sub-functions arising during the composition process. Intuitively, functional reconfiguration maintains consistency throughout the decomposition process. This consistency contributes even further to interpretability and explainability because it ensures that each part of the model adheres to a uniform structure or set of principles. 

We analyze both functional decomposition and functional reconfiguration through the lenses of parameterized complexity theory. In our framework, functions are represented symbolically, using ordered binary decision diagrams (OBDDs), a prominent formalism for the symbolic representation of Boolean functions \cite{bryant1992symbolic}. The reconfiguration process is described using a Boolean circuit $\circuit$. Finally, the function class $\mathcal{F}$ is specified using a second-order finite automaton $\functionclassautomaton$, an automata-theoretic formalism for the representation of classes of functions. Our main results state that functional reconfiguration, and functional decomposition can be solved in fixed-parameter linear time when parameterized by the number of factor functions, the width of the OBDDs representing these functions, the size of the circuit $C$ and the size of the second-order automaton $\functionclassautomaton$.

\subsection{Our Results}
\label{subsection:OurResults}

For a fixed Boolean circuit $C$ with $k$ inputs and a fixed class of functions $\mathcal{F}$, we consider the problem of decomposing a given Boolean function $f:\{0,1\}^{n}\rightarrow \{0,1\}$ as a $C$-combination of functions $f_1,\dots,f_k:\{0,1\}^n\rightarrow \{0,1\}$ from $\mathcal{F}$. More specifically, we require that for each $x\in \{0,1\}^n$, $f(x) = \gatefunction_{\circuit}(f_1(x),\dots,f_k(x))$, where $\gatefunction_{\circuit}$ is the $k$-bit function computed by $C$. We say that the sequence $f_1,\dots,f_k$ is a $(C,\mathcal{F})$-decomposition of $f$. 

In order for the $(C,\mathcal{F})$-decomposition problem to make sense from a computational point of view, we also need to specify a model for the representation of the input function $f$ and factor functions $f_1,\dots,f_k$, as well as a model for the representation of the class of functions $\mathcal{F}$. In our framework, we represent Boolean functions symbolically, using ordered binary decision diagrams (OBDDs), which are one of the most prominent formalisms for symbolic computation \cite{bryant1992symbolic}. 
An OBDD $D$ is, essentially, an acyclic deterministic finite automaton where the set of states is partitioned into a sequence of levels $X_0X_1...X_n$ and the set of transitions is partitioned into a 
sequence of layers $\B_1\B_2\dots \B_n$, where each layer $\B_i$ contains only transitions from states in $X_{i-1}$ to $X_{i}$. Such an OBDD $D$ represents a function $f_D:\{0,1\}^n\rightarrow \{0,1\}$ where for each $x\in \{0,1\}^n$, $f(x) = 1$ if and only if $D$ accepts $x$. At the same time that OBDDs may provide a much more concise representation of a Boolean function when compared to representations based on truth tables or decision trees, OBDDs enjoy a series of algorithmic properties, such as efficient computation of union, intersection, complementation, etc \cite{bryant1992symbolic}. 

An important complexity measure in the context of symbolic computation is the {\em width} of an OBDD, that is to say, the maximum number of states in one of its levels  \cite{Bollig2014width,Bollig2012symbolic,Hachtel1993symbolic,Woelfel2006symbolic,Sawitzki2004implicit}.
For each fixed $\solutionWidth\in \N$, an OBDD of width at most $\solutionWidth$ can be encoded as a word over the alphabet $\detCalB(\solutionWidth)$ of all layers of width at most $\solutionWidth$.
This allows one to define classes of Boolean functions using automata theoretic formalisms. More specifically, a finite automaton $\functionclassautomaton$ over $\detCalB(\solutionWidth)$ defines the class
$\mathcal{F}(\functionclassautomaton) = \{f_\D\;:\; \D\in \Lang(\functionclassautomaton)\}$ of all functions computed 
by OBDDs accepted by $\functionclassautomaton$. A crucial feature of this approach is the fact that, as shown in \cite{MeloOliveira2022}, second-order finite automata can be canonized with respect to the {\em class of functions} it represents, and classes of functions represented by second-order automata are closed under 
usual Boolean operations, as well as several other higher-order operations \cite{MeloOliveira2022}. 

In this work, we show that the $(C,\mathcal{F}(\functionclassautomaton))$-decomposition problem can be solved in fixed-parameter tractable linear time when parameterized by the number $\ncomponents$ of factor functions, the width $\solutionWidth$ of the OBDDs representing the factor functions $\ffunction_1,\dots,\ffunction_k$, the number $m$ of gates in the circuit $\circuit$, and the size $|\functionclassautomaton|$ of the second-order finite automaton representing the class $\mathcal{F}$.  

\parproblem{\pnFunctionDecomposition}{
Numbers $\ncomponents,\solutionWidth,m\in \N$, an OBDD $\D$, a Boolean circuit $\circuit$ with $\ncomponents$ inputs and $m$ gates, and a second-order finite automaton $\functionclassautomaton$ over $\detCalB(\solutionWidth)$.}{$\ncomponents$, $\solutionWidth$, $m$, $|\functionclassautomaton|$.}{Are there OBDDs $\D_1,\dots,\D_{\ncomponents}\in \mathcal{F}(\functionclassautomaton)$ of width at most 
$\solutionWidth$ such that $\ffunction_{\D}=\decomposerFunction_{\circuit}(\ffunction_{\D_1},\dots,\ffunction_{\D_{\ncomponents}})$?}

Our main result (Theorem \ref{theorem:UpperBoundsGenFunction}) states that 
\pnFunctionDecomposition can be solved in time 
$$2^{p^{O(2^\circuitsize)}}\cdot |\functionclassautomaton|^\ncomponents \cdot |\D|$$ where $|\D|$ is the size of the input OBDD representing the function to be decomposed. We note that when the parameters $\ncomponents$, $\solutionWidth$, $m$, $|\functionclassautomaton|$ are fixed, our algorithm works in linear time on the size of the input OBDD $\D$. 

Let us give a brief motivation for our parameterization. First, we note that if an OBDD of length $n$ has width $\solutionWidth$ then the number of bits necessary to specify the OBDD is $O(\solutionWidth\cdot n)$. In usual decomposition problems, the parameter $k$ is often set to $2$, since the goal in this case is to factorize a given mathematical object into two simpler mathematical objects. Any Boolean function on $k$ inputs can be computed by a Boolean circuit with at most $O(2^k/k)$ gates \cite{shannon1949synthesis}. Therefore the parameter $m$ is upper bounded by a function of $k$. Nevertheless, it is also reasonable to have the size of the circuit as a parameter because this size can be much smaller than the trivial upper bound.
For instance, the AND function on $k$ bits can be computed by a circuit with 
$k-1$ gates (of fan-in 2). Finally, many interesting classes of functions, can be represented by second-order finite automata of constant size \cite{Kuske21,MeloOliveira2022}.

We prove our main result within the framework of functional reconfiguration. More specifically, we introduce the notion of {\em reconfiguration width} of a circuit $C$ with 
respect to input functions $f_1,\dots,f_k$ (Definition \ref{definition:ReconfigurationWidth}), a complexity measure that may be of independent interest. Intuitively, this measure captures the maximum complexity of a
function arising during the process of reconfiguring the input functions $f_1,\dots,f_k$
into the output function $f=\gatefunction_{\circuit}(f_1,\dots,f_k)$ according to the specification provided by circuit $C$.
In Section \ref{section:ReconfigurationWidth}, we introduce the problem \pnRecFunctionDecomposition\!\!\!, a reconfiguration variant 
of \pnFunctionDecomposition where the reconfiguration width is requested to be below a 
threshold $\width$. We show that this problem can be solved in time $2^{\mathcal{O}(\circuitsize \cdot \width^2\log\width)}\cdot |\functionclassautomaton|^\ncomponents \cdot |\D|$ (Theorem \ref{theorem:MainTechnicalDecomposition}). Our main result follows by combining Theorem \ref{theorem:MainTechnicalDecomposition} with an upper bound on 
the reconfiguration width in terms of the width of the factor functions and of the size of the circuit (Lemma \ref{lemma:TransformationComplexityUpperBound}). 
 

%% file: 02-Preliminaries.tex
\section{Preliminaries} 
\label{section:preliminaires}

\subsection{Basics}
We let $\N \eqdef  \{0,1, \ldots \}$ be the set of {\em natural numbers}, 
and $\Nplus\eqdef \N \setminus \{0\}$ be the set of positive natural numbers. 
For each $n \in \N$ we set $[n] \eqdef \{1,2, \ldots,n\}$. According to this notation,
$[0] \eqdef \emptyset$. For each $\width \in \Nplus$, we let $\dbset{\width} \eqdef \{0,1,\ldots,
\width-1\}$. 

An \emph{alphabet} is any finite, non-empty set $\alphabet$. A \emph{string}
over $\alphabet$ is a finite sequence of symbols from $\alphabet$.  We 
let $\alphabet^*$ be the set of all strings over $\alphabet$, including the empty
string $\emptystring$, and let $\alphabet^+ =
\alphabet^*\setminus\{\emptystring\}$ be the set of non-empty strings over
$\alphabet$. A \emph{language} over $\alphabet$ is any subset $\languageL
\subseteq \alphabet^*$. For each $k\in \Nplus$, we let $\alphabet^k$ be the
set of all strings of length $k$ over $\alphabet$.

\subsection{$\alpha$-Regular Languages} 
The proofs of our results heavily rely on techniques from automata theory. For this reason, we briefly recall some basic concepts about finite automata and regular languages. 

A \emph{finite automaton} is a tuple $\automaton =
(\alphabet,\states,\initialStates,\finalStates,\transitions)$ where $\alphabet$
is an alphabet, $\states$ is a finite set of {\em states}, $\initialStates\subseteq \states$ is a set of {\em initial states}, 
$\finalStates\subseteq \states$ is a set of {\em final states}, and $\transitions
\subseteq \states\times \alphabet \times \states$ is a set of {\em transitions}.
A string $\stringx \in \alphabet^{\lengthString}$ is said to be \emph{accepted}
by $\automaton$ if there is a sequence
$$
(\stateq_0,\stringx_1,\stateq_1)(\stateq_1,\stringx_2,\stateq_2)\dots
(\stateq_{n-1},\stringx_n,\stateq_n)
$$ 
such that $\stateq_0\in \initialStates$,
$\stateq_n\in \finalStates$, and  for each $i\in [n]$,
$(\stateq_{i-1},\stringx_i,\stateq_i)$ is a transition in $\transitions$. The
\emph{language} of $\automaton$ is defined as the set 
$$
\Lang(\automaton) =
\{\stringx\;|\; \stringx \mbox{ is accepted by $\automaton$}\}
$$
of all strings
accepted by $\automaton$. We note that finite automata considered in this work 
may be non-deterministic, meaning that for some state $\stateq$ and some symbol 
$a\in \alphabet$, we may have two states $\stateq'$ and $\stateq''$ such that 
$(\stateq,a,\stateq')$ and $(\stateq,a,\stateq'')$ are transitions in $\automaton$. 

\begin{definition} \label{definition:Regularity} Let $\regularityalpha\in \N$
and $\languageL \subseteq \alphabet^*$. We say that $\languageL$ is
$\regularityalpha$-regular if there is a finite
automaton $\automaton$ over $\alphabet$ with at most $\regularityalpha$ states
such that $\Lang(\automaton) = \languageL$.  \end{definition}

In the product construction for finite automata, two automata \( \automaton \) and \( \automaton' \) are combined into a single automaton \(\automaton\times \automaton'\) whose states are pairs of states \((q, q')\) from \( \automaton \) and \( \automaton' \), respectively, with transitions defined such that \((q, q') \xrightarrow{\sigma} (r, r')\) if and only if \(q \xrightarrow{a} r\) in \( A \) and \(q' \xrightarrow{a} r'\) in \( \automaton' \). Initial (final) states are pairs of initial (final) states of the respective automata. It can be shown that $\languageL(\automaton\times \automaton') = \languageL(\automaton)\cap \languageL(\automaton')$ \cite{hopcroft1971n}. The following observation is a direct consequence of this construction.  

\begin{observation}
\label{observation:ProductRegularities}
Let $\languageL$ be an $\alpha$-regular language, and $\languageL'$ be an $\alpha'$-regular language. Then $\languageL\cap \languageL'$ is $(\alpha\cdot \alpha')$-regular. 
\end{observation}

\subsection{Boolean Circuits}
\label{subsection:BooleanCircuits}

We view a Boolean circuit with input variables 
$\{\variable_1,\dots,\variable_{\ncomponents}\}$ as a sequence 
$\circuit= (\gate_1,\gate_2,\dots,\gate_{m})$ of gates
satisfying the following conditions: 
\begin{enumerate} 
\item Each gate $\gate_i$ is of one of the following types: 
\begin{enumerate}
\item a variable in the set $\{\variable_1,...,\variable_{\ncomponents}\}$; 
\item a \texttt{NOT} gate $\gate_i = \texttt{NOT}(\gate_j)$ for some $j< i$; 
\item an \texttt{AND} gate $\gate_i = \texttt{AND}(\gate_j,\gate_l)$ for some $j,l< i$;
\item an \texttt{OR} gate $\gate_i = \texttt{OR}(\gate_j,\gate_l)$ for some $j,l<i$.
\end{enumerate}
\item For each $i\in [\ncomponents]$, there is exactly one $j\in [m]$ such 
that $\gate_j = \variable_i$.  
\item $g_m$ is the only gate that is an output gate, i.e., a gate
that is not an input to any other gate. 
\end{enumerate}

Given a Boolean circuit $\circuit = (\gate_1,\gate_2,\dots,\gate_{m})$ and an index $i\in [m]$, we let $\gatefunction_{\gate_i}:\{0,1\}^k\rightarrow \{0,1\}$ denote the Boolean function computed at gate $\gate_i$. We may write $\gatefunction_{\circuit}$ to denote the Boolean function $\gatefunction_{\gate_m}$ computed by the output gate of $\circuit$. 
We say that a function $\ffunction:\{0,1\}^{\lengthString}\rightarrow \{0,1\}$ 
is a Boolean combination of functions $\ffunction_1,\dots,\ffunction_{\ncomponents}:\{0,1\}^{\lengthString}\rightarrow \{0,1\}$ if there is a Boolean 
circuit $\circuit$ such that 
$\ffunction = \decomposerFunction_{\circuit}(\ffunction_1,\dots,\ffunction_{\ncomponents})$. 
More specifically, for each $\stringx\in \{0,1\}^{\lengthString}$, we have 
that $\ffunction(\stringx) = \decomposerFunction_{\circuit}(\ffunction_1(\stringx),\dots,\ffunction_{\ncomponents}(\stringx))$. 

Let $\circuit = (\gate_1,\dots,\gate_{\ngates})$ be a Boolean circuit. The depth of a gate $\gate_i$ is inductively defined as follows. 
If $\gate_i$ is an input gate, then $\depth(\gate_i) = 0$. On the other hand, if $\gate_i$ is a non-input gate, then the depth of $\gate_i$ is 
defined as the maximum depth of an input for $\gate_i$ plus $1$. 

%% file: 03-AutomataVsFunctionClasses.tex
\section{Automata vs Function Classes}

In its most traditional setting, finite automata are used to represent sets of strings over a given finite alphabet $\alphabet$, or equivalently, as a way to represent functions of type $\alphabet^*\rightarrow \{0,1\}$. More recently, in a wide variety of contexts, automata theoretic formalisms have been defined with the intention to represent sets of sets of strings, or equivalently, function classes \cite{jain2012learnability,Case2013,abu2017advice,Kuske21,MeloOliveira2022}. In this section we describe the approach from \cite{MeloOliveira2022}, which is based on the notion of a second-order finite automaton. Our interest in this formalism stems from the fact that function classes represented in this way enjoy several useful closure and decidability properties, such as closure under intersection, union, bounded-width complementation, besides having decidable inclusion and emptiness of intersection tests. 

\subsection{Ordered Binary Decision Diagrams}
\label{subsection:OBDDs}

We start by defining the notion of an ordered binary decision diagram using a slightly 
different notation than the one usually employed in the
literature \cite{bryant1992symbolic}. The reason is that 
in our work, it will be more convenient to view an ordered binary decision diagram of width at most 
$\width$ as a sequence of symbols over an alphabet $\detCalB(\width)$ of $\width$-layers. 

Let $\width\in \Nplus$. We call a subset $\B\subseteq \dbset{\width}\times \{0,1\} \times \dbset{\width}$ a {\em $\width$-layer}. We let $\detCalB(\width)$ denote the set of all 
$\width$-layers. For each
$B\in \detCalB(\width)$, we let 
$$\Dom(\B) \eqdef \{\stateq \;|\; \exists (a,\stateq'),
(\stateq,a,\stateq') \in \B\}$$
be the \emph{domain} of $\B$, and
$$\im(\B) \eqdef
\{\stateq' \;|\; \exists (\stateq,a), (\stateq,a,\stateq') \in \B\}$$
be the
\emph{image} of $\B$. 
Given $\lengthString,\width\in \Nplus$, an {\em ordered binary decision diagram} (OBDD)
of length $\lengthString$ and width at most $\width$ is a sequence 
$\D = \B_1\B_2\dots \B_{\lengthString}$
of $\width$-layers satisfying the following conditions: 

\begin{enumerate}
\item $\Dom(\B_1) = \{0\}$,
\item for each $i\in [n-1]$, $\im(\B_i)=\Dom(\B_{i+1})$,
\item for each $i\in [n]$, each $\stateq\in \Dom(\B_i)$, and each $a\in \{0,1\}$,
there is a unique $\stateq'$ such that $(\stateq,a,\stateq')\in \B_i$. 
\end{enumerate}

We say that a string $\stringx=\stringx_1 \stringx_2 \ldots \stringx_{\lengthString}$ is accepted by $\D$ if there is a sequence 
$$(\stateq_0,
\stringx_1,\stateq_1)(\stateq_1,\stringx_2,\stateq_2)\ldots(\stateq_{\lengthString-1},\stringx_{\lengthString},\stateq_{\lengthString})$$
where for each $i \in [\lengthString]$, $(\stateq_{i-1},\stringx_i,\stateq_i)
\in \B_i$ , and $\stateq_{\lengthString} \in \dbset{\width}\setminus\{0\}$. Intuitively, 
$\stringx$ is accepted by $\D$ if it reaches a nonzero element in the image of the last layer of $\D$. 
The language of $\D$ is the set 
$$
\Lang(\D)= \{\s\; | \; \s \text{ is accepted by } \D\}.
$$
We may also regard an OBDD of length $\lengthString$ as a representation of a binary 
function $f_{\D}:\{0,1\}^{\lengthString}\rightarrow \{0,1\}$. More specifically, for 
each $s\in \{0,1\}^n$, we set $f_{\D}(s)=1$ if and only if $s\in \Lang(\D)$. In other words, OBDD $\D$ represents {\em support set} of $f_{\D}$, that is, the set $\ffunction^{-1}(1)$ of all strings in $\{0,1\}^n$
that $\ffunction$ sends to $1$.

\begin{figure}[t]
\centering
\includegraphics[]{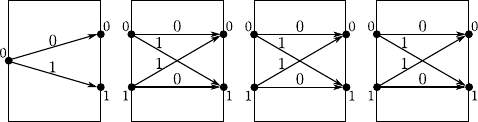}
\caption{An OBDD of width $2$ and length $4$ accepting all strings in $\{0,1\}^4$ of odd parity. Note that each string of odd parity (say, $1011$) reaches the element $1$ in the image of the last layer, while each string of even parity (say, $1010$) reaches the element $0$ in the image of the last layer.}
\label{figure:OBDDExample}
\end{figure}

In Figure \ref{figure:OBDDExample}, we depict an example of an OBDD. Each layer is 
represented by a box, while each triple $(\stateq,a,\stateq')$ in such a layer is
represented by an arrow $\stateq\xrightarrow{a}\stateq'$.

We say that an OBDD $\D = \B_1\B_2,\dots\B_{\lengthString}$ is {\em
normalized} if for each $i\in [\lengthString]$, and each two $j,j'\in
\im(\B_i)$ with $j<j'$, the lexicographically-first string in $\{0,1\}^i$
that reaches $j$ is smaller than the lexicographically-first string in
$\{0,1\}^{i}$ that reaches $j'$. 
For instance, the OBDD $\D$ depicted in Figure \ref{figure:OBDDExample} is normalized because for each $i\in [4]$, 
the lexicographically-first string that reaches the element $0$
in the image of $\B_i$ is the string $0^i$, while the lexicographically-first
string that that reaches the element $1$ is the string $0^{i-1}1$. 

The size of an OBDD $\D = \B_1\B_2\dots \B_n$, denoted by $|\D|$ 
is defined as $|\D| = |\Dom(\B_1)|+\sum_{i\in [n]} |\im(B_i)|$. The width 
of $\D$ is defined as $\ODDwidth(\D) = \max\{|\im(\B_i)|\}$. 
The next theorem, which is well known in the OBDD literature, states that each finite language $\languageL\subseteq \{0,1\}^n$
corresponds to a unique normalized OBDD $\D$ of minimum size such that 
$\Lang(\D) = \languageL$. Additionally this OBDD has also minimum width. 

\begin{theorem}[\cite{bryant1992symbolic,MeloOliveira2022}] \label{theorem:CanonicalForm} 
For each OBDD $\D$ there is a unique normalized OBDD $\can(\D)$ of
minimum size with the property that $\Lang(\D)=\Lang(\can(\D))$.
Additionally, $\can(\D)$ has minimum width among all 
OBDDs $\D'$ with $\Lang(\D') = \Lang(\D)$.
\end{theorem}

We note that if $\D$ is an OBDD of length $\lengthString$ and width at most 
$\width$ then $\can(\D)$ can be constructed in time $O(\width\cdot \lengthString)$,
using the standard algorithm for minimization of acyclic finite automata, followed 
by normalization \cite{bryant1992symbolic,MeloOliveira2022}.
The fact that $\can(\D)$ is normalized guarantees that 
it is {\em syntactically unique}, meaning that for each two OBDDs $\D$ and $\D'$, 
$\Lang(\D) = \Lang(\D')$ if and only if $\can(\D)=\can(\D')$. In most applications requiring minimal OBDDs, it is enough to consider OBDDs that are minimal up to renaming of states. In our applications however, syntactic uniqueness is an important requirement.

\subsection{Second Order Finite Automata}
\label{subsection:RegularFunctionClasses}

Recall that we view an OBDD $\D$ of width $\width$ and length $\lengthString$ as 
a sequence $\D = \B_1\B_2\dots\B_n$ of $\width$-layers. In other words, $\D$ is simply
a string of length $\lengthString$ over the alphabet $\detCalB(\width)$ of all layers. 
We note that not every string in the set $\detCalB(\width)^n$ is a valid OBDD
of length $\lengthString$ and width $\width$, since not all such strings satisfy
Conditions $1$-$3$ stated in Subsection \ref{subsection:OBDDs}. We let 
$\detCalB(\width)^{\circ \lengthString}$ denote the set of all strings in 
$\detCalB(\width)^{\lengthString}$ that {\em are} OBDDs. 
We let $\detCalB(\width)^{\circledast} = \bigcup_{\lengthString\in \N} \detCalB(\width)^{\circ \lengthString}$ 
denote the set containing the empty string, and all OBDDs of width at most $\width$.
The following proposition states that the set $\detCalB(\width)^{\circledast}$ 
is $2^{\width}$-regular. 

\begin{proposition}
\label{proposition:RegularityOBDDs}
For each $\width\in \N$, there is a finite automaton $\functionclassautomaton(\width)$
with $2^{\width}$ states such that 
$\Lang(\functionclassautomaton(\width)) =  \detCalB(\width)^{\circledast}$. 
\end{proposition}
\begin{proof}
    The states of $\functionclassautomaton(\width)$ are the subsets of $\dbset{\width}$. The only initial state is the state $\{0\}$. All states are final. For each pair of states $S$ and $S'$, and each $B\in\detCalB(\width)$ there is a transition $(S,B,S')$ if and only if $S = \Dom(B)$ and $S' = \im(B)$. It is straightforward to check that a sequence $D = B_1B_2...B_n$ is accepted by $\functionclassautomaton(\width)$ if and only if $D\in \detCalB(\width)^{\circ \lengthString}$. 
\end{proof}

Since each OBDD $\D$ of width at most $\width$ represents a function $\ffunction_{\D}$, a set of OBDDs of width at most $\width$ may be seen as a representation of a function class. That is to say, the class consisting of all functions associated with OBDDs in the set. We will be particularly concerned with function classes that can be finitely represented using finite automata over the alphabet $\detCalB(\width)$. 

\begin{definition}
\label{definition:SecondOrderFA}
We say that a finite automaton $\functionclassautomaton$ over
$\detCalB(\width)$ is a {\em second-order finite automaton} if
$\Lang(\functionclassautomaton)\subseteq \detCalB(\width)^{\circledast}$. We let 
$$\classfunctionsF(\functionclassautomaton) = \{\ffunction_{\D}\;:\; \D\in \Lang(\functionclassautomaton)\}$$ be the function class represented by $\functionclassautomaton$. 
\end{definition}

We say that a function class $\classfunctionsG$ is {\em decisional} if there is some $\width\in \N$ and some second-order finite automaton $\functionclassautomaton$ over $\detCalB(\width)$ such that $\classfunctionsG = \classfunctionsF(\functionclassautomaton)$. 

Let $\width,\width'\in \N$, and let $\classfunctionsG\subseteq \classfunctionsF(\functionclassautomaton(\width))$. We define the width-$\width$ complement of $\classfunctionsG$ as the function class $\overline{\classfunctionsG}^{\width} = \classfunctionsF(\functionclassautomaton(\width))\backslash \classfunctionsG$. 
The next theorem, from \cite{MeloOliveira2022}, states some basic closure and decidability properties for decisional function classes. 
We note that these properties do not follow directly from usual closure and decidability properties of regular languages. The problem is that several OBDDs of width at most $\width$ may represent the same function. Therefore, it is not in general possible to establish a one-to-one correspondence between functions in $\classfunctionsF(\functionclassautomaton)$ and OBDDs in $\Lang(\functionclassautomaton)$.

\begin{theorem}[\cite{MeloOliveira2022}]
\label{theorem:MeloOliveira}
Decinsional function classes are closed under union, intersection, and bounded-width complementation. Furthermore, inclusion and intersection emptiness for regular function classes are decidable.
\end{theorem}

%% file: 04-FunctionalReconfiguration.tex
\section{Functional Reconfiguration}
\label{section:ReconfigurationWidth}

A Boolean circuit $\circuit$ with $\ncomponents$ inputs may be regarded as the specification of a 
process that {\em reconfigures} input functions 
$\ffunction_1,\dots,\ffunction_{\ncomponents}$ into the output 
function $\ffunction = \gatefunction_{\circuit}(\ffunction_1,\dots,\ffunction_{\ncomponents})$, where 
$\gatefunction_{\circuit}$ is the $k$-bit function computed by $C$.
In this section, we introduce a suitable measure of complexity for such a reconfiguration process.
More specifically, we let the {\em reconfiguration width} of $C$ with respect to $\ffunction_1,\dots,\ffunction_{\ncomponents}$ be the 
maximum complexity (i.e. width) of a function $\gatefunction_{\gate_i}(\ffunction_1,\dots,\ffunction_{\ncomponents})$ 
arising during the reconfiguration process. 

Given a function $\ffunction:\{0,1\}^{\lengthString}\rightarrow \{0,1\}$ we let 
$\canonicalOBDD(\ffunction)$ denote the (unique) canonical OBDD $\D$ of 
length $\lengthString$ such that $\ffunction_{\D} = \ffunction$ (see Theorem \ref{theorem:CanonicalForm}). In other words, 
the canonical OBDD $\D$ with language $\Lang(\D) = f^{-1}(1)$.  
We note that $\D$ has minimum width among all OBDDs representing $\ffunction$.
Using this notion, we can define the notion of {\em reconfiguration width} of a $\ncomponents$-input 
Boolean circuit $\circuit$ with respect to functions $\ffunction_1,\dots,\ffunction_{\ncomponents}:\{0,1\}^{n}\rightarrow \{0,1\}$. 

\begin{definition}[Reconfiguration Width]
\label{definition:ReconfigurationWidth}
Let $\circuit = (\gate_1,\dots,\gate_m)$ be a Boolean circuit with $\ncomponents$ inputs, and 
$\ffunction_1,\dots,\ffunction_{\ncomponents}:\{0,1\}^{\lengthString}\rightarrow \{0,1\}$ be Boolean functions. 
The reconfiguration width of $\circuit$ with respect to 
$\ffunction_1,\dots,\ffunction_{\ncomponents}$ is defined as 
$$\widthtransformation(\circuit,\ffunction_1,\dots,\ffunction_{\ncomponents}) = \max_{i\in [m]}  \ODDwidth(\canonicalOBDD(\gatefunction_{\gate_i}(\ffunction_1,\dots,\ffunction_{\ncomponents}))).$$ 
\end{definition}

The next problem is a refinement of \pnFunctionDecomposition. In this refinement,
we require that the reconfiguration width of $\circuit$ with respect 
to $\ffunction_{\D_1},\dots,\ffunction_{\D_{\ncomponents}}$ is bounded by $\width$. 
In other words, one wishes to decompose a function $\ffunction_{\D}$ into
simpler functions $\ffunction_{\D_1},\dots,\ffunction_{\D_{\ncomponents}}$ in such a 
way that all functions occurring in the process of reconfiguring 
$\ffunction_{\D_1},\dots,\ffunction_{\D_{\ncomponents}}$  into 
$\ffunction_{\D}$ have low complexity (i.e. width at most $\width$). 

\parproblem{\pnRecFunctionDecomposition}{
Numbers $\ncomponents,\width,\solutionWidth, m \in \N$ with $\solutionWidth<\width$, an OBDD $\D$, a Boolean circuit $\circuit$ with $\ncomponents$ inputs and $m$ gates, and a 
second-order finite automaton
$\functionclassautomaton$ over $\detCalB(\solutionWidth)$.}{$\ncomponents$, $\width$, $m$,$|\functionclassautomaton|$.}{Are there OBDDs $\D_1,\dots,\D_{\ncomponents}\in \mathcal{F}(\functionclassautomaton)$ of width at most 
$\solutionWidth$ such that
$\ffunction_{\D}=\decomposerFunction_{\circuit}(\ffunction_{\D_1},\dots,\ffunction_{\D_{\ncomponents}})$
and
$\widthtransformation(\circuit,\ffunction_{\D_1},\dots,\ffunction_{\D_{\ncomponents}})\leq
\width$?}

The next theorem, whose proof is detailed in section \ref{section:ProofMainTheorem},
states that \pnRecFunctionDecomposition is fixed-parameter tractable when parameterized
by the number of factor functions $\ncomponents$, the reconfiguration width 
$\width$, the size $m$ of the circuit $C$, and the size of the second-order 
automaton representing the class of functions $|\functionclassautomaton|$. We note that 
the parameter $\circuitsize$ is upper bounded by $O(2^{\ncomponents}/\ncomponents)$ since 
any Boolean function on $\ncomponents$ inputs can be computed by a circuit of size at most 
$O(2^{\ncomponents}/\ncomponents)$\cite{shannon1949synthesis}. Additionally, the dependency on the parameter $|\functionclassautomaton|$ is of the form $|\functionclassautomaton|^{k}$. Therefore when $k$ and $w$ are fixed, our algorithm for \pnRecFunctionDecomposition runs in polynomial time even if the number of states of 
$\functionclassautomaton$ is polynomial in $n$. 

\begin{theorem}
\label{theorem:MainTechnicalDecomposition}
\pnRecFunctionDecomposition can be solved in time 
$$2^{\mathcal{O}(\circuitsize \cdot \width^2\log\width)}\cdot |\functionclassautomaton|^\ncomponents \cdot |\D|.$$
\end{theorem}

In the problem \pnRecFunctionDecomposition, 
we need to 
specify the maximum width of an OBDD occurring during the reconfiguration process. 
Next, we estimate an upper bound for reconfiguration width 
in terms of the maximum width $p$ of the canonical OBDD representing one of the factor functions and the depth $d$ of the decomposer circuit $C$. 
By noting that $d$ is upper bounded by the number of gates $m$ of the circuit, and that $m$ is upper bounded by $O(2^k/k)$ we also obtain bounds for solving  
\pnFunctionDecomposition parameterized by $\solutionWidth$ and $m$ or by $\solutionWidth$ and $\ncomponents$. 
 
\begin{lemma}
\label{lemma:TransformationComplexityUpperBound}
Let $\D_1,\ldots,\D_{\ncomponents}$ be OBDDs in 
$\detCalB(\solutionWidth)^{\circ \lengthString}$, and let 
$\circuit$ be a circuit with $\ncomponents$ inputs and 
depth at most $\circuitDepth$. Then 
$\widthtransformation(\circuit,\ffunction_{\D_1},\dots,\ffunction_{\D_{\ncomponents}})\leq \solutionWidth^{2^\circuitDepth}$. 
\end{lemma}
\begin{proof}
Let $\circuit = (\gate_1,\gate_2,\dots,\gate_{\circuitsize})$. 
We claim that for each $i\in [\circuitsize]$, the width of the OBDD 
$\canonicalOBDD(\gatefunction_{\gate_i}(\ffunction_{\D_1},\dots,\ffunction_{\D_\ncomponents}))$ is upper-bounded by $\solutionWidth^{2^\circuitDepth}$, 
where $\circuitDepth$ is the depth of gate $\gate_i$ in $\circuit$. 
The proof is by induction on the depth of a gate. In the base case, let $\gate_i$ be a gate of depth $0$, that is to say, an input gate. Then $\gatefunction_{\gate_i}(\ffunction_{\D_1},\dots,\ffunction_{\D_\ncomponents}) = \ffunction_{\D_j}$ for 
some $j\in [k]$. Since, $\D_j \in \detCalB(\solutionWidth)^{\circ \lengthString}$, we have that the width of $\canonicalOBDD(\gatefunction_{\gate_i}(\ffunction_{\D_1},\dots,\ffunction_{\D\ncomponents}))$ is at most 
$\solutionWidth^{2^0}=\solutionWidth$. Now, let $\gate_i = \texttt{AND}(\gate_j,\gate_k)$ be a gate of depth $\depth$, and suppose that the statement of the lemma holds for every gate of depth at most $\depth-1$. Then by the induction hypothesis, there are OBDDs 
$\D_{g_j},\D_{g_k} \in \detCalB(\solutionWidth^{2^{\circuitDepth-1}})$ computing the functions $\gatefunction_{g_j}(\ffunction_{\D_1},\dots,\ffunction_{\D\ncomponents})$ and $\gatefunction_{g_k}(\ffunction_{\D_1},\dots,\ffunction_{\D_\ncomponents})$ respectively. 
Since $\gate_j = \texttt{AND}(\gate_j,\gate_k)$, we have that $\ffunction_{\D_{g_i}} = \ffunction_{\D_{g_j}}\wedge \ffunction_{\D_{g_k}}$, or alternatively, $\Lang(\D_{\gate_i}) = \Lang(\D_{\gate_j})\cap \Lang(\D_{\gate_k})$. 
It can be shown (see Lemma 15 of \cite{MeloOliveira2022}) that given OBDDs $\D$ and $\D'$ of width at most $\solutionWidth$ there is 
an OBDD of width at most $\solutionWidth^2$ accepting the language $\Lang(\D)\cap \Lang(\D')$. Therefore, we have that there is an OBDD of width at most $(\solutionWidth^{2^{d-1}})^2 = \solutionWidth^{2^d}$ computing the function $\gatefunction_{g_i}(\ffunction_{D_1},\dots,\ffunction_{\D_{\ncomponents}})$.
A similar argument can be proved in the case that $\gate_i = \texttt{OR}(\gate_i,\gate_j)$. 
\end{proof}

By plugging Lemma \ref{lemma:TransformationComplexityUpperBound} into 
Theorem \ref{theorem:MainTechnicalDecomposition} we get the following upper bounds for solving 
\pnFunctionDecomposition. 

\begin{theorem}
\label{theorem:UpperBoundsGenFunction}
\pnFunctionDecomposition can be solved in time
\begin{enumerate}
\item $2^{p^{O(2^d)}}\cdot |\functionclassautomaton|^\ncomponents \cdot |\D|$, 
\item $2^{p^{O(2^\circuitsize)}}\cdot |\functionclassautomaton|^\ncomponents \cdot |\D|$, 
\item $2^{p^{2^{O(2^{\ncomponents}/k)}}}\cdot |\functionclassautomaton|^\ncomponents \cdot |\D|.$
\end{enumerate}
\end{theorem}
\begin{proof}
Item 1 follows directly by plugging Lemma \ref{lemma:TransformationComplexityUpperBound} into
Theorem  \ref{theorem:MainTechnicalDecomposition}. Item 2 follows by noting that a circuit of size $m$ has depth at most $m$. Item $3$ follows from the fact that any Boolean function with $k$ inputs can be computed by a circuit of size $O(2^k/k)$. 
\end{proof}

%% file: ProofOfMainTheorem.tex
\section{Proof of Theorem \ref{theorem:MainTechnicalDecomposition}}

\label{section:ProofMainTheorem}

In this section, we prove our main technical theorem (Theorem \ref{theorem:MainTechnicalDecomposition}). Our proof is based on automata theoretic techniques. 
In particular, we will use automata to define relations, and to operate with these relations. We recall some basic concepts in this area for completeness. Further literature in this subject can be found for instance in \cite{MeloOliveira2022,blumensath2000automatic}. 

\subsection{Relations as Languages}



Given a set $S$ and a positive number $\arityvalue \in \Nplus$, we let
$S^{\times \arityvalue}$ denote the set of all $\arityvalue$-tuples of elements
from $S$. Given sets $X$ and $Y$ we denote by $Y^X$ the set of all functions
of the form $f:X\rightarrow Y$.

Let $\stringx_1,\stringx_2,\dots,\stringx_{\ncomponents}$ be strings 
of length $\lengthString$ over alphabets $\alphabet_1, \alphabet_2,\dots,\alphabet_{\arityvalue}$
respectively. The \emph{tensor product} of $\stringx_1,\stringx_2,\dots,\stringx_{\arityvalue}$
is the string $$\stringx_1\otimes \stringx_2\otimes \dots \otimes \stringx_{\arityvalue}$$
of length $\lengthString$ over the alphabet 
$\alphabet_1\times \dots \times \alphabet_{\arityvalue}$ 
where for each $i\in [n]$, the symbol at position $i$ is the $\arityvalue$-tuple  
$(\stringx_{1,i},\stringx_{2,i},\dots,\stringx_{\arityvalue,i})$. 

%




Let $\lengthString\in \Nplus$, and $\relation \subseteq \alphabet_1^n\times \dots \times
\alphabet_{r}^\lengthString$ be an $\arityvalue$-ary relation consisting of $\arityvalue$-tuples
of strings over the alphabets $\alphabet_1,\dots,\alphabet_{\arityvalue}$, respectively. 
The language associated with
$\relation$ is the set $\Lang(\relation)\subseteq (\alphabet_1 \times
\ldots \times \alphabet_r)^+$ defined as follows. 
$$ \Lang(\relation) = \{\stringx_1\otimes \stringx_2\otimes \dots \otimes
\stringx_{\arityvalue} \;:\;
(\stringx_1,\stringx_2,\dots,\stringx_{\arityvalue})\in \relation\}.  $$

We say that $\relation$ is $\regularityalpha$-regular, for some
$\regularityalpha\in \Nplus$, if the language $\Lang(\relation)$ is
$\regularityalpha$-regular. 

\subsection{Selectors}

Let $\Sigma$ be an alphabet, and $r,m\in \Nplus$.
A $(r,m)$-selector is a pair 
$(\relation,\selectorTuple)$ where $\relation\subseteq (\Sigma^*)^{\times r}$ is a 
relation of arity $r$ and 
$\selectorTuple = (i_1,...,i_r)$ is a tuple in $[m]^{\times r}$. 
%
%
%
The language of $(\relation,\selectorTuple)$ is defined as 
$$\selectorlanguage{\relation}{\selectorTuple} = \{u_1\otimes \dots \otimes u_m \in (\Sigma^{\times m})^+ \;|\; u_{i_1} \otimes \dots \otimes u_{i_r}\in \Lang(\relation)\}.$$ 


\begin{proposition}\label{label:selectorLanguageRegularity}
Let $(\relation,\selectorTuple)$ be a $(r,m)$-selector. If $\relation$ is an $\regularityalpha$-regular relation, then the language $\selectorlanguage{\relation}{\selectorTuple}$ is $\regularityalpha$-regular.  
\end{proposition}
\begin{proof}
Let $\selectorTuple=(i_1,\dots,i_r)$ and consider the map $\mathfrak{m}_{\selectorTuple}:\Sigma^{\times m}\rightarrow \Sigma^{\times r}$ where $\mathfrak{m}_\selectorTuple(a_1,\dots,a_m)=(a_{i_1},\dots, a_{i_r})$.
Then, we have that $\selectorlanguage{\relation}{\selectorTuple}=\mathfrak{m}_\selectorTuple^{-1}(\Lang(\relation))$. In other words, $\selectorlanguage{\relation}{\selectorTuple}$ is the inverse homomorphism of $\Lang(\relation)$
under the map $\mathfrak{m}$. It is well known from automata theory that inverse homomorphisms derived from maps between alphabets preserve number of states in an automaton. More specifically, in our case, given an automaton
$A$ with at most $\regularityalpha$ states accepting the language $\Lang(\relation)$, we consider the automaton $A'$ obtained from $A$ by replacing each transition 
$(q,(a_1,\dots,a_r),q')$ with the set of transitions $\{(q,b,q')\;:\;b\in \mathfrak{m}^{-1}(a_1,\dots,a_r)\}$. Then, one can readily check that $\Lang(A') = \mathfrak{m}_\selectorTuple^{-1}(\Lang(\relation))$. 
This shows that $\selectorlanguage{\relation}{\selectorTuple}$ is $\regularityalpha$-regular. 
\end{proof}

Given a language $\languageL\subseteq (\Sigma^{\times m})^*$ and a $(r,m)$-selector $(\relation,\selectorTuple)$ where $\selectorTuple=(i_1,\dots,i_r)$, we let 
$$\selection{\languageL}{\relation}{\selectorTuple} = \{u_1\otimes u_2 \otimes \dots \otimes u_m \in \languageL\;|\; u_{i_1}\otimes \dots \otimes u_{i_r} \in \Lang(\relation)\}$$ 
be the subset of $\languageL$ whose strings in entries $(i_1,\dots,i_r)$ are consistent with relation $\relation$.
We call $\selection{\languageL}{\relation}{\selectorTuple}$ the $(\relation,\selectorTuple)$-selection of $\languageL$. 

\begin{proposition}
\label{proposition:SelectionLemma}
Let $(\relation,\selectorTuple)$ be a $(r,m)$-selector and 
$\languageL \subseteq (\Sigma^{\times m})^*$. If $\languageL$ is $\regularityalpha$-regular and $\relation$ is $\regularitybeta$-regular, then $\selection{\languageL}{\relation}{\selectorTuple}$ is $(\regularityalpha\cdot \regularitybeta )$-regular. 
\end{proposition}
\begin{proof}
The language $\selection{\languageL}{\relation}{\selectorTuple}$ is the intersection of $\languageL$ with the language $\selectorlanguage{\relation}{\selectorTuple}$ which by Proposition \ref{label:selectorLanguageRegularity} is $\beta$-regular since $R$ is $\beta$-regular. 
Given an automaton $\automaton$ with at most $\regularityalpha$ states accepting $\languageL$, and an automaton $\automaton'$ with at most $\regularitybeta$ states accepting $\Lang(\relation)$, the 
standard product construction in automata theory yields an automaton $\automaton''$ with at most $\regularityalpha\cdot \regularitybeta$ states accepting $\selection{\languageL}{\relation}{\selectorTuple}$. 
%
\end{proof}

\subsection{Useful Relations}
In this section we give upper bounds on the regularity of relations representing standard set theoretic operations over second-order funite automata. 

The following theorem is an adaptation of Theorem 9 of \cite{MeloOliveira2022} to
the context of OBDDs. It states that the relation, $\relation^{det}_{can}$, pairing an OBDD with its deterministic canonical OBDD is $2^{\mathcal{O}(\width \cdot \log \width)}$-regular.

\begin{theorem}
\label{theorem:RegularityCanonical}[\cite{MeloOliveira2022}]
Let $\widthone,\widthtwo \in \N$, and $\width$ be equal 
to $\max\{\widthone,\widthtwo\}$.
The following relation is $2^{\mathcal{O}(\width \cdot \log \width)}$ regular.
$$
\begin{array}{l}
\relation^{det}_{can}(\widthone,\widthtwo) =  
\{(\D,\can(\D))\;|\; \D\in \detCalB(\widthone)^\circledast, \can(\D)\in \detCalB(\widthtwo)^\circledast\}.
\end{array}
$$
\end{theorem}

Next, we define several useful relations. 

\begin{definition}
\label{definition:Relations}
Below, we let $\width_1,\width_2,\width_3\in \N$. 
\begin{enumerate}
\setlength{\itemsep}{0.3em}
\item $\relation_{=}(\widthone,\widthtwo) =$  $\{(\D_1,\D_2)\;|\; \D_i\in 
\detCalB(\width_i)^\circledast, \Lang(\D_1) = \Lang(\D_2)\}$. 
\item $\relation_{\cap}(\widthone,\widthtwo,\widththree) =$ $\{(\D_1,\D_2,\D_3)  \;|\;    D_i \in \detCalB(\width_i)^\circledast,  \Lang(\D_3) = \Lang(\D_1) \cap \Lang(\D_2)\}$. 
\item $\relation_{\cup}(\widthone,\widthtwo,\widththree) =$  $\{(\D_1,\D_2,\D_3)\;|\; \D_i\in \detCalB(\width_i)^\circledast, \Lang(\D_3) = \Lang(\D_1) \cup \Lang(\D_2)\}$. 
\item $\relation_{\neg}(\widthone,\widthtwo) =$ $\{(\D_1,\D_2) \;|\; \D_i \in \detCalB(\width_i)^\circledast, \Lang(\D_1) = \overline{\Lang(\D_2)} \}$. 
\end{enumerate}
\end{definition}

Using Theorem \ref{theorem:RegularityCanonical}, one can derive the following corollary. 

\begin{corollary}
\label{corollary:RegularityRelations}
Let  $\widthone,\widthtwo,\widththree\in \N$, and 
$\width$ be equal to $\max\{\widthone,\widthtwo\}$ (or $\max\{\widthone,\widthtwo,\widththree\}$ where appropriate).
\begin{enumerate}
\item $\relation_{=}(\widthone,\widthtwo)$ is $2^{\mathcal{O}(\width \cdot \log\width)}$-regular. 
\item $\relation_{\cap}(\widthone,\widthtwo,\widththree)$ is $2^{\mathcal{O}(\width^2\log\width)}$-regular.  
\item $\relation_{\cup}(\widthone,\widthtwo,\widththree)$ is $2^{\mathcal{O}(\width^2\log\width)}$-regular. 
\item $\relation_{\neg}(\widthone,\widthtwo)$ is  $2^{\mathcal{O}(\width \cdot \log\width)}$-regular
\end{enumerate}
\end{corollary}

The proof of Corollary \ref{corollary:RegularityRelations} is given i appendix \ref{appedixOmitted}. In what follows, we will write $\relation_{=}(\width)$ to mean $\relation_{=}(\widthone,\widthtwo)$ when $\widthone = \widthtwo$. This notation extends analogously to the other relations in the above corollary.

\subsection{Main Development}

Given functions
$\ffunction_1:\{0,1\}^{n}\rightarrow \{0,1\}$ and
$\ffunction_2:\{0,1\}^{n}\rightarrow \{0,1\}$, we let $f_1\wedge
\ffunction_2:\{0,1\}^{n}\rightarrow \{0,1\}$ be the function whose support is
the intersection of the supports of $\ffunction_1$ and $\ffunction_2$, 
$f_1\vee \ffunction_2:\{0,1\}^{n}\rightarrow \{0,1\}$ be the function 
whose support is the union of the supports of $\ffunction_1$ and $\ffunction_2$, 
and $\neg \ffunction_1:\{0,1\}^n\rightarrow \{0,1\}$ be the function 
whose support is the complement of the support of $\ffunction_1$.

\begin{definition}[Consistency]\label{Definition:Consistency}
Let $\circuit = (\gate_1,\gate_2,\dots,\gate_{m})$ be a Boolean circuit with $\ncomponents$ inputs $\{\variable_1,\dots,\variable_{\ncomponents}\}$, and let $\D_1 \otimes \D_2 \otimes \ldots \otimes \D_m \in (\detCalB(\w)^{\times m})^{\circledast}$. 
For each $\ell\in [m]$, we say that $\D_1\otimes \D_2\otimes \dots \otimes \D_m$ is \emph{consistent with $\circuit$} up to the $\ell$-th gate if the following conditions are satisfied for each  $i\in [\ell]$: 
\begin{enumerate}
    \item if $\gate_i = \texttt{AND}(\gate_j,\gate_h)$ then $\ffunction_{\D_i} = \ffunction_{\D_j} \wedge \ffunction_{\D_h}$, 
    \item if $\gate_i = \texttt{OR}(\gate_j, \gate_h)$ then $\ffunction_{\D_i} = \ffunction_{\D_j} \vee \ffunction_{\D_h}$,
    \item if $\gate_i = \texttt{NOT}(\gate_j)$ then $\ffunction_{\D_i} = \neg \ffunction_{\D_j}$. 
\end{enumerate}
If $\D_1\otimes \D_2 \otimes \dots \otimes \D_m$ is consistent with $\circuit$ up to gate $m$, then we simply say that $\D_1\otimes \D_2 \otimes \dots \otimes \D_m$ is consistent with $\circuit$. 
\end{definition}

Let $m\in \Nplus$, and $(\detCalB(\width)^{\times m})^{\circledast}$ be the language over the alphabet 
$\detCalB(\width)^{\times m}$ consisting of all 
strings of the form $\D_1\otimes \D_2\otimes \dots \otimes \D_{m}$
where for each $i\in [m]$, $\D_i$ is an OBDD in $\detCalB(\w)^{\circledast}$, and $|\D_1|=|\D_2|=\dots=|\D_m|$. 

The next proposition (taken from \cite{MeloOliveira2019} Proposition 9) upper bounds the regularity of tensor products of regular languages.
\begin{proposition}[\cite{MeloOliveira2019}] 
\label{proposition:TensorProduct}
Let  $r_1,r_2\in \Nplus$. Let $L_1$ be a language in 
$(\Sigma_1^{\times r_1})^+$, and $L_2$ be a language in $(\Sigma_2^{\times r_2})^{+}$.
If $L_1$ is $\alpha_1$ regular and $L_2$ is $\alpha_2$-regular, then the language
$L_1\otimes L_2 \in (\{\Sigma_1 \cup \Sigma_2\}^{\times(r_1+r_2)})^{+}$ is $\alpha_1\cdot \alpha_2$-regular. 
\end{proposition}


Combining the above result with Proposition \ref{proposition:RegularityOBDDs}, we get the following corollary.

\begin{corollary}
\label{proposition:RegularityProduct}
For each $\w,m\in \Nplus$, the language $(\detCalB(\w)^{\times m})^{\circledast}$ is $2^{m\cdot \w}$-regular. 
\end{corollary}
\begin{proof}
There exists an automata accepting language $\detCalB(\w)^{\circledast}$ with $2^{\w}$ many states by Proposition \ref{proposition:RegularityOBDDs}. Applying \ref{proposition:TensorProduct} gives regularity $2^{m\cdot \w}$.
\end{proof}

The following lemma establishes an upper bound on the regularity of 
the language of words of the form, 
$\D_1\otimes \D_2\otimes \dots \otimes \D_m$, that 
are consistent with a given circuit $\circuit$. 

\begin{lemma}\label{mOBDDsRegularity}
Let $\circuit = (\gate_1,\gate_2,\dots,\gate_{\ngates})$ be a circuit on $\ncomponents$ inputs. The language,
$$
\begin{array}{l}
\consistencyLanguage(\width,\circuit) = \{\D_1 \otimes \D_2 \otimes \ldots \otimes \D_m \in  (\detCalB(\w)^{\times m})^{\circledast} \;|\; 
\D_1 \otimes \D_2 \otimes \ldots \otimes \D_m \;\mbox{consistent with} \; \mathcal{\circuit} \}
\end{array}
$$ 
is $2^{\mathcal{O}(m \cdot \width^2\log\width)}$-regular.
\end{lemma}
\begin{proof}

Let $i \in [m]$ and assume that the gates of $\circuit$ are ordered in some appropriate topological ordering. To each internal $\gate_i$, 
we can associate a $(r_i,m)$-selector $(\relation_i,\selectorTuple_i)$ where $r_i$ is the arity of the gate plus one.
More specifically, 
\begin{equation*}
(\relation_i,\selectorTuple_i) =
\left\{
\begin{array}{ll}
(\relation_\cap(\w),(j,h,i)) & \mbox{ if } \gate_i = \texttt{AND}(\gate_j,\gate_h), \\[4pt] 
(\relation_\cup(\w),(j,h,i)) & \mbox{ if } \gate_i = \texttt{OR}(\gate_j,\gate_h), \\[4pt] 
(\relation_\neg(\w),(j,i))   & \mbox{ if } \gate_i = \texttt{NOT}(\gate_j).
\end{array}
\right.
\end{equation*}

For each $\ell \in [m]$, we let $\consistencyLanguage_{\ell}(\width,\circuit)$ be the language consisting of all 
strings $\D_1\otimes\D_2\otimes\ldots\otimes\D_m$ in $(\detCalB(\w)^{\times m})^{\circledast}$ 
that are consistent with $\circuit$ up to gate $\ell$.

It turns out that the language  $\consistencyLanguage_{\ell}(\width,\circuit)$ can be defined inductively using 
selectors. More specifically, let $\languageL_1,\languageL_2,\dots,\languageL_m \subseteq (\detCalB(\w)^{\times m})^{\circledast}$
be the sequence of languages inductively defined as follows. 
\begin{enumerate}   
    \item $\languageL_1 = (\detCalB(\width)^{\times m})^{\circledast}$ 
    \item If $\gate_i$ is an input gate, then $\languageL_i=\languageL_{i-1}$
    \item If $\gate_i$ is an internal gate then $\languageL_i = \selection{\languageL_{i-1}}{\relation_i}{\selectorTuple_i}$. 
\end{enumerate}

We claim that for each $i\in [m]$, the language $\languageL_i$ is equal to the language $\consistencyLanguage_i(\width,\circuit)$ and that $\consistencyLanguage_i(\width,\circuit)$ has regularity 
$2^{\mathcal{O}(i \cdot (\width^2\log\width) +m\width)}$.

The proof is by induction on $i$. In the base case, we have that $\languageL_1 = 
(\detCalB(\w)^{\times m})^{\circledast}$ since 
by definition of circuit, $\gate_1$ is necessarily an input gate. Therefore $\consistencyLanguage_1(\width,\circuit) = \languageL_1$. Since $\languageL_1$ has complexity $2^{m\cdot \width} < 2^{\mathcal{O}((\width^2\log\width) +m\width)}$, from  Corollary \ref{proposition:RegularityProduct}, the claim is proved in the base case.

Now, let $i\in [m-1]$, and assume that the claim holds for $i$. We show that the claim also holds for $i+1$. 

If $\gate_{i+1}$ is an input gate then $\languageL_{i+1} = \languageL_{i}$. Since by assumption, $\languageL_i$ is 
$2^{\mathcal{O}(i \cdot (\width^2\log\width) +m\width)}$-regular, we have that $\languageL_{i+1}$ is  
$2^{\mathcal{O}((i+1) \cdot (\width^2\log\width) +m\width)}$-regular. 

If $\gate_{i+1}$ is an internal node, then $\languageL_{i+1} = \selection{\languageL_{i}}{\relation_{i+1}}{\selectorTuple_{i+1}}$. By the selection lemma, we take the product of the regularity of $\relation_{i+1}$, which can be at most $2^{\mathcal{O}( \width^2\log\width)}$, and the regularity of $\languageL_{i}$. This results in regularity $2^{\mathcal{O}(i+1 \cdot ( \width^2\log\width) +m\width)}$ for $\languageL_{i+1}$. Meanwhile, $\languageL_{i+1} = \consistencyLanguage_{i+1}(\width,\circuit)$ since $\languageL_{i+1} = \selection{\languageL_{i}}{\relation_{i+1}}{\selectorTuple_{i+1}}$ making it consistent with $\circuit$ up to $\gate_{i+1}$. 

Having now proven our claim, see that $\consistencyLanguage_m( \width,\circuit) = \consistencyLanguage(\width,\circuit)$ with regularity $2^{\mathcal{O}(m \cdot \width^2\log\width)}$.

\end{proof}

By definition of $\consistencyLanguage(\width,\circuit)$, it can be shown that the reconfiguration width of $\circuit$ with respect to OBDDs appearing in some tuple of $\consistencyLanguage(\width,\circuit)$ at entries corresponding to the input gates of $\circuit$, is bounded by $\width$.
More formally, the following corollary observes an upper bound on the reconfiguration width of $\circuit$ with respect to functions 
$\ffunction_{\D_{j_1}},\dots,\ffunction_{\D_{j_{\ncomponents}}}$ where for each 
$i\in [\ncomponents]$, $g_{j_i}$ is the gate corresponding to the $i$-th input.

\begin{corollary}\label{lemma:transformationWidth}
Let $\circuit = (\gate_1,\gate_2,\dots,\gate_{m})$ be a circuit on $k$ inputs with input gates $\gate_{j_1}=\variable_1,\gate_{j_2}=\variable_2,\dots,\gate_{j_{\ncomponents}} = \variable_{\ncomponents}$. For any $\D_1 \otimes \D_2 \otimes \ldots \otimes \D_m \in \consistencyLanguage(\width,\circuit)$, $\widthtransformation(\circuit,\ffunction_{\D_{j_1}},\dots,\ffunction_{\D_{j_\ncomponents}}) \leq \w$.
\end{corollary}

\begin{proof}
For each $i\in [m]$, 
$\gatefunction_{\gate_i}(\ffunction_{\D_{j_1}},\dots,\ffunction_{\D_{j_k}}) = \ffunction_{\D_i}$ by the definition of consistency (Definition \ref{Definition:Consistency}). Moreover, since 
for each $i\in [m]$, $\D_i \in \detCalB(\w)^{\circ n}$ we have that 
$\ODDwidth(\canonicalOBDD(\gatefunction_{\gate_i}(\ffunction_{\D_{j_1}},\dots,\ffunction_{\D_{j_k}}))) = \ODDwidth(\canonicalOBDD(\ffunction_{\D_i})) \leq w$ (recall that the canonical form of an OBDD $\D$ is the minimum deterministic OBDD accepting the language of $\D$).
\end{proof}

The next proposition observes a straightforward upper bound on the regularity of a second order finite automata whose language is a single OBDD. 

\begin{proposition}\label{RegularityOfD}
Let $\D = \B_1\B_2\ldots \B_n$ be an OBDD in $\detCalB(\w)^{\circledast}$. The language $\languageL_\D = \{\D\}$ is accepted by an automaton with $n+1$ states.
\end{proposition}
\begin{proof}
Let $A$ be the automaton with set of states $\{q_0, q_1, \ldots q_n\}$, initial state $q_0$, final state $q_n$, and set of transitions $\{(q_{i-1},\B_i,q_i)| i \in \dbset{n+1} \}$. One can verify that $A$ accepts the language $\languageL_\D$.
\end{proof}


By retaining sets of OBDDs appearing together in some tuple of $\consistencyLanguage(\width,\circuit)$ at entries corresponding to the input gates of $\circuit$ we can express the language of tuples of OBDDs whose boolean combination under $C$ is a non-zero function (whose support set is not empty). For any chosen OBDD $\D$, one can further restrict this language to tuples of OBDDs representing functions whose boolean combination is $\ffunction_\D$.

We let $\solutionLanguage(\width,\D,\circuit)$ be the set of all strings
$\D_1 \otimes \D_2 \otimes \ldots \otimes \D_k \in (\detCalB(\width)^{\times k})^{\circ n}$ 
in $(\detCalB(\width)^{\times k})^{\circ n}$ 
with the property that $\ffunction_\D = \gatefunction_\circuit(\ffunction_{\D_1}, \ldots,\ffunction_{\D_k})$ and $\widthtransformation(\circuit,\ffunction_{\D_1},\dots,\ffunction_{\D_\ncomponents}) \leq \w$.

\begin{lemma}\label{proposition:kOBDDsRegularity}
Let $\circuit = (\gate_1,\dots,\gate_{m})$ be a circuit with $\ncomponents$ input variables $\{\variable_1, \variable_2, \ldots, \variable_{\ncomponents}\}$, and let $\D$  be an OBDD in 
$\detCalB(\w)^{\circ n}$. Then $\solutionLanguage(\width,\D,\circuit)$ is $2^{\mathcal{O}(m\cdot \width^2\log\width)}\cdot n$-regular.
\end{lemma}
\begin{proof}

Let $\languageL= \consistencyLanguage(\width,\circuit) \otimes \languageL_\D$ be the tensor product of language $\consistencyLanguage(\width,\circuit)$ with the language $\languageL_\D$. We note that this is the language of all tensor products of $m+1$ OBDDs where the tensor product of the first $m$ OBDDs belongs $\consistencyLanguage(\width,\circuit)$, and the $(m\!+\!1)$-th OBDD is $\D$. 
Let $\gate_{j_1}=\variable_1,\dots, \gate_{j_{\ncomponents}} = \variable_{\ncomponents}$ be the input gates of $\circuit$. Then by Lemma \ref{lemma:transformationWidth}, we have that $\widthtransformation(\circuit,\ffunction_{\D_{j_1}},\dots,\ffunction_{\D_{j_\ncomponents}}) \leq \w$. 

By Lemma \ref{mOBDDsRegularity}, $\consistencyLanguage(\width,\circuit)$ is $2^{\mathcal{O}(m \cdot \width^2\log\width)}$-regular and by Proposition \ref{RegularityOfD} language $\languageL_\D$ is $n+1$-regular. Therefore, $\languageL$ is $2^{\mathcal{O}(m \cdot \width^2\log\width)}\cdot n$-regular Proposition \ref{proposition:TensorProduct}. Let $\relation= \relation_=(\w)$ and $\selectorTuple=(m,m+1)$. Since $\relation_=(\w)$ is $2^{\mathcal{O}( \width \cdot \log\width)}$-regular, the $(\relation,\selectorTuple)$-selection, $\selection{\languageL}{\relation}{\selectorTuple}$, is $2^{\mathcal{O}(m \cdot \width^2\log\width)}\cdot n$-regular by Lemma \ref{proposition:SelectionLemma}. 

 We note that 
 $\selection{\languageL}{\relation}{\selectorTuple}$ 
  is the language of all tensor products of $m+1$ OBDDs where the tensor product of the first $m$ OBDDs belongs $\consistencyLanguage(\width,\circuit)$, and $\gatefunction_{\circuit}(\ffunction_{j_1},\dots,\ffunction_{j_{\ncomponents}}) = \ffunction_{\D}$.
 
 Now, let $\pi:\detCalB(\w)^{\times m+1} \rightarrow \detCalB(\w)^{\times k}$ be the map where $\pi(\B_1,\B_2,\ldots,\B_{m+1})=(\B_{j_1},\B_{j_2},\ldots,\B_{j_{\ncomponents}})$. Intuitively $\pi$ deletes all coordinates in the tuple $(\B_1,\B_2,\ldots,\B_{m+1})$ except for the coordinates $j_1, j_2, \ldots, j_{\ncomponents}$. Note that in the resulting tuple, for each $i\in [\ncomponents]$, $\B_{j_i}$ is the layer occurring in the coordinate corresponding to variable $x_i$. Since the regularity of a language is preserved under mappings between alphabets, we have that $\pi(\selection{\languageL}{\relation}{\selectorTuple})$ is also 
$2^{\mathcal{O}(m \cdot \width^2\log\width)}\cdot n$-regular.
We note that $\pi(\selection{\languageL}{\relation}{\selectorTuple})$
is the language of all $\D_1 \otimes \D_2 \otimes \ldots \otimes \D_\ncomponents$ in 
$(\detCalB(\width)^{\times k})^{\circ n}$ such that 
$\ffunction_\D$ is a $\gatefunction_\circuit$-combination of $\ffunction_{\D_1}, \ldots,\ffunction_{\D_k}$ and $\widthtransformation(\circuit,\ffunction_{\D_1},\dots,\ffunction_{\D_\ncomponents}) \leq \w$. This proves the lemma. 
\end{proof}




Finally, we may further restrict OBDDs appearing in tuples of $\solutionLanguage(\width,\D,\circuit)$ to represent functions belonging to predetermined function classes.

Let $\functionclassautomaton = (\functionclassautomaton_1,\dots,\functionclassautomaton_{\ncomponents})$ 
be a tuple of $\alpha$-regular second-order finite automata. 
We define $\solutionLanguage_{\functionclassautomaton}(\width,\D,\circuit)$ be the subset of $\solutionLanguage(\width,\D,\circuit)$ where for each $\D_1 \otimes \D_2 \otimes \ldots \otimes \D_k \in \solutionLanguage(\width,\D,\circuit)$, $\D_1 \in \Lang(\functionclassautomaton_1),\dots,\D_\ncomponents \in \Lang(\functionclassautomaton_{\ncomponents})$,

$$
\solutionLanguage_{\functionclassautomaton}(\width,\D,\circuit)= \{\D_1 \otimes \D_2 \otimes \ldots \otimes \D_k \in \solutionLanguage(\width,\D,\circuit) \;|\;  \D_i \in \Lang(\functionclassautomaton_i), i \in [\ncomponents]\}
$$

Theorem \ref{theorem:MainTechnicalDecomposition} is a direct consequence
of the following theorem.

\begin{theorem}\label{theorem:generalFunctionDecomposition}
Let $\circuit = (\gate_1,\dots,\gate_{m})$ be a circuit with $\ncomponents$ input variables $\{\variable_1, \variable_2, \ldots, \variable_{\ncomponents}\}$ and depth at most $\circuitDepth$. 
Let $\D$ be a given OBDD in $\detCalB(\w)^{\circ n}$, and let $\functionclassautomaton = (\functionclassautomaton_1,\dots,\functionclassautomaton_{\ncomponents})$ be a tuple of 
$\regularityalpha$-regular second-order finite automata. Then, the language $\solutionLanguage_{\functionclassautomaton}(\width,\D,\circuit)$ is $2^{\mathcal{O}(m \cdot \width^2\log\width)}\cdot \regularityalpha^\ncomponents \cdot n$-regular.
\end{theorem}
\begin{proof}
 Since second-order finite automata $\functionclassautomaton_1,\dots,\functionclassautomaton_{\ncomponents}$ are $\regularityalpha$-regular, there exists 
 unary relations
 $\relation_1, \ldots,\relation_k$, $\relation_i \subseteq \bigcup_{\lengthString\in \Nplus }\detCalB(\w_i)^{\circ \lengthString}$, for $i \in [k]$, such that $\classfunctionsF(\functionclassautomaton_i) = \{\ffunction_{\D}\;:\; \D\in \relation_i\}$. Let $\languageL=\{\D_1 \otimes \D_2 \otimes \ldots \otimes \D_k | \D_i \in \relation_i\}$ be the language of tensor products of $k$ OBDDs $\D_1 \otimes \D_2 \otimes \ldots \otimes \D_k$ such that $\D_1 \in \relation_1, \ldots\D_\ncomponents \in \relation_\ncomponents$. The language $\languageL$ is $\regularityalpha^k$-regular by Corollary \ref{proposition:RegularityProduct}.
 
 We can then take the intersection of $\languageL$ and $\solutionLanguage(\width,\D,\circuit)$. Language $\solutionLanguage(\width,\D,\circuit)$ is $2^{\mathcal{O}(m \cdot \width^2\log\width)}\cdot n$-regular by Lemma \ref{proposition:kOBDDsRegularity}
 . Taking the product construction of automata accepting $\languageL$ and $\solutionLanguage(\width,\D,\circuit)$ we get that $\languageL \cap \solutionLanguage(\width,\D,\circuit)$ is $2^{\mathcal{O}(m \cdot \width^2\log\width)}\cdot \regularityalpha^\ncomponents \cdot n$-regular
 .
 We note that $\languageL\cap \solutionLanguage(\width,\D,\circuit) = \solutionLanguage_{\functionclassautomaton}(\width,\D,\circuit)$ as it is the subset of
 $\solutionLanguage(\width,\D,\circuit)$ where for each
$\D_1 \otimes \D_2 \otimes \ldots \otimes \D_k \in \solutionLanguage(\width,\D,\circuit)$, it is the case that $\ffunction_{\D_1} \in \classfunctionsF(\functionclassautomaton_1),\dots,\ffunction_{\D_\ncomponents} \in \classfunctionsF(\functionclassautomaton_{\ncomponents})$.
\end{proof}

%% file: 05-Applications.tex
\section{Applications}

In this section we describe some prominent applications for the \pnFunctionDecomposition problem. The first application is concerned with the notion of a generalized $k$-junta. That is a function that is a Boolean combination of $k$ functions belonging to some class $\mathcal{F}$. The crucial difference is that here, the decomposer circuit $C$ is not given a priori. In the second application we obtain an FPT algorithm for the OBDD factorization problem, parameterized by the number of factor OBDDs and the width of the OBDDs. This is problem is analogous to the DFA factorization problem \cite{DBLP:conf/concur/JeckerM021,jecker2020unary,Kupferman2013}, which does not admit and FPT algorithm parameterized by the number of factor automata. 

\subsection{Generalized Juntas}
\label{section:Juntas}

For $\lengthString,i\in \N$, with $i\leq \lengthString$, the $(\lengthString,i)$-hypercube 
is the function $H_{\lengthString,i}:\{0,1\}^{\lengthString}\rightarrow \{0,1\}$ 
such that for each $x\in \{0,1\}^n$, $H_{\lengthString,i}(x)=1$ if and only if $x_i=1$. 
A function $f:\{0,1\}^n\rightarrow \{0,1\}$ is said to be a $\ncomponents$-junta if there are 
numbers  $j_1,\dots,j_k$ such that $f$ is a Boolean combination of $H_{\lengthString,j_1},\dots,H_{\lengthString,j_k}$. Intuitively, $f$ is a $k$-junta if there are coordinates $j_1,\dots,j_k \in [n]$ such that for each input $x\in \{0,1\}^n$, the value $f(x)$ is determined by the values of $x_{j_1},\dots,x_{j_k}$. 
The problem of determining whether a function is a $k$-junta has been studied under a wide variety of contexts \cite{blais2010testing}. It turns out that when the input function $f$ is specified by an OBDD $\D$, then for each $k\in \N$, there is a simple algorithm that decides whether $f$ is a $k$ junta in time linear on the size of $\D$.

\begin{observation}
Given an OBDD $\D$ of length $n$ and a number $k\in \N$, one can determine whether the function 
$f_{\D}:\{0,1\}^n\rightarrow \{0,1\}$ is a $k$-junta in time linear on the size of $\D$. 
\end{observation}
\begin{proof}
 We assume that $\D$ is minimized. Otherwise, we just apply the standard minimization algorithm for OBDDs. 
 Let $i\in [n]$ and suppose that $f_\D$ does not depend on the value of $x_i$. Then, every transition from any state $q$ of layer $i$ to layer $i+1$ must be of the form $(q,0,q')$ and $(q,1,q')$. In other words, 
 setting $x_i$ to $1$ or $0$ from state $q$ should both lead to state $q'$. If this were not the case, that is, if there were transition $(q,0,q')$ and $(q,1,q'')$ in layer $i$ with $q'\neq q''$, then $q'$ and $q''$ could be merged since they accept the same words (because $f_D$ does not depend on $x_i$). This would be in contradiction with the assumption that $D$ is minimized. Therefore, given a minimized OBDD, one can just traverse each layer checking for each layer $i$ whether it contains transitions $(q,0,q')$ and $(q,1,q'')$ for $q'\neq q''$. Doing so, we identify the set of variables $Y \subseteq {x_1,\dots,x_n}$ on which $D$ depends. Deciding whether $D$ is a $k$-Junta hence boils down to checking whether $|Y| \leq k$.
\end{proof}

More general notions of juntas have been considered in the literature under a variety of contexts \cite{de2021robust,de2019your}. In these contexts, instead of requiring the $k$ factor functions $f_1,\dots,f_k$ to be hypercubes, we require that these functions belong to some well-behaved class of functions $\mathcal{F}$. 

\begin{definition}
\label{definition:GeneralizedJunta }
Let $\mathcal{F}$ be a class of Boolean functions and $k\in \mathbb{N}$. We say that a function $f:\{0,1\}^n\rightarrow \{0,1\}$ is a $(k,\mathcal{F})$-junta if there are $k$ functions $f_1,\dots,f_k:\{0,1\}^n\rightarrow \{0,1\}$ in $\mathcal{F}$ such that $f$ is a Boolean combination of $f_1,\dots,f_k$. 
\end{definition}

Note that equivalently,  $f:\{0,1\}^n\rightarrow \{0,1\}$ is a $(k,\mathcal{F})$-junta, if it admits a $(\circuit,\mathcal{F})$-decomposition $f_1,\dots,f_k:\{0,1\}^n\rightarrow \{0,1\}$ for some circuit $C$. This motivates the definition of a variant of \pnFunctionDecomposition where instead of providing specifying a composition circuit $\circuit$ we simply specify an upper bound on the size of such a circuit. We call this generalization \pnGenJunta.  

\parproblem{\pnGenJunta}{
Numbers $\ncomponents,\solutionWidth,m\in \N$, an OBDD $\D$, and a second-order finite automaton $\functionclassautomaton$ over $\detCalB(\solutionWidth)$.}{$\ncomponents$, $\solutionWidth$, $m$, $|\functionclassautomaton|$.}{Is there a Boolean circuit $\circuit$ with $\ncomponents$ inputs and at most $m$ gates, and OBDDs $\D_1,\dots,\D_{\ncomponents}\in \functionclassautomaton$ of width at most $\solutionWidth$  
such that $\ffunction_{\D} =
\decomposerFunction_{\circuit}(\ffunction_{\D_1},\dots,\ffunction_{\D_{\ncomponents}})$?}

The next theorem states that \pnGenJunta can be solved within the same asymptotic bounds provided in Theorem \ref{theorem:UpperBoundsGenFunction}. 

\begin{theorem}
\label{corollary:Juntas}
\pnGenJunta can be solved in time
\begin{enumerate}
\item $2^{p^{O(2^d)}}\cdot |\functionclassautomaton|^\ncomponents \cdot |\D|$, 
\item $2^{p^{O(2^\circuitsize)}}\cdot |\functionclassautomaton|^\ncomponents \cdot |\D|$, 
\item $2^{p^{2^{O(2^{\ncomponents}/\ncomponents)}}}\cdot |\functionclassautomaton|^\ncomponents \cdot |\D|.$
\end{enumerate}
\end{theorem}
\begin{proof}
The theorem follows by enumerating all possible circuits of size at most $m$ and then applying Theorem 
\ref{theorem:UpperBoundsGenFunction}. Given that a circuit with $m$ gates can be specified with $O(m\log m)$ 
bits, we have that the total number of such circuits is $2^{O(m\log m)}$. Given that 
$2^{O(m\log m)}\cdot 2^{p^{O(2^\circuitsize)}}\cdot |\functionclassautomaton|^\ncomponents \cdot |\D|$
is still $2^{p^{O(2^\circuitsize)}}\cdot |\functionclassautomaton|^\ncomponents \cdot |\D|$, 
we have 
proved Item 2. If we want to parameterize the problem only by $\ncomponents$ and $\solutionWidth$, then we 
just need to enumerate all circuits with up to $m=O(2^k/k)$ gates. Therefore, we have Item 3. On the other 
hand, if we want to parameterize only by the depth $d$ of the circuit then we note that any circuit of depth 
$d$ has at most $2^{O(d)}$ gates. Therefore, the complexity parameterized by depth is 
$2^{2^{O(d)}}\cdot 2^{p^{O(2^d)}}\cdot |\functionclassautomaton|^\ncomponents \cdot |\D|$ which is 
still asymptotically $2^{p^{O(2^d)}}\cdot |\functionclassautomaton|^\ncomponents \cdot |\D|$. Therefore, we have Item 1. 
\end{proof}

\subsection{OBDD Factorization}
\label{OBDDFactorization}

A prominent factorization problem in the context of automata theory is the {\em DFA factorization problem}  \cite{DBLP:conf/concur/JeckerM021,jecker2020unary,Kupferman2013}. In this problem we are given a DFA $A$ with $w$ states and the goal is to determine whether there are DFAs $A_1,\dots,A_k$ each with at most $w-1$ states such that 
$\languageL(A) = \languageL(A_1)\cap \dots \cap \languageL(A_{\ncomponents})$.

An analogous problem in the context of OBDDs, the OBDD factorization problem, we are given an OBDD of width at most $w$ and the goal is to determine whether there exist OBDDs $D_1,\dots,D_k$ each of width at most $w-1$ such that $\languageL(\D) = \languageL(D_1)\cap \dots\cap \languageL(D_k)$. 
A direct consequence of our main theorem is that this problem can be solved in fixed parameter linear time (that is, linear in $|D|$) by instantiating the second-order automaton $\functionclassautomaton$ as the automaton that accepts all OBDDs of width at most $w-1$.

\begin{corollary}
\label{corollorary:OBDDFactorization}
Given an OBDD $\D$ of length $n$ and width at most $w$, one can determine in time $2^{w^{O(2^k)}}\cdot |\D|$, 
 whether there are OBDDS $\D_1,\dots,D_k$ of width at most $w-1$ such that $\languageL(\D) = \languageL(D_1)\cap \dots\cap \languageL(D_k)$. 
\end{corollary}

%% file: 06-ConclusionAndRelatedWork.tex
\section{Conclusion and Related Work}

In this work we introduced a new framework to address functional reconfiguration and functional decomposition problems by combining techniques from automata theory and parameterized complexity theory. In our results, we assume that the function $f:\{0,1\}^n\rightarrow \{0,1\}$ is specified explicitly at the input using an OBDD $\D$. We note that this assumption can be relaxed if instead of an OBDD we are given access to an oracle that is able to answer membership and equivalence queries with respect to the language $L = \{x\in \{0,1\}^n\;:\; f(x)=1\}$. More specifically, in the traditional model of learning with membership and equivalence queries, the objective is to learn a 
regular language $L$ over a given alphabet $\Sigma$, where the learner interacts with an oracle (a minimally adequate teacher) 
capable of answering two types of queries:

\begin{enumerate}
\item {\em Membership queries}: the learner selects a word in $x \in \Sigma^{*}$ and the teacher replies whether or not $x \in L$.
\item {\em Equivalence queries}: the learner selects an hypothesis automaton $H$ and the teacher replies whether or not
$L$ is the language of $H$. If this is not the case, the teacher provides the learner with a counter-example word $x\in \Sigma^*\backslash L$.
\end{enumerate}

A classic result due to Angluin \cite{angluin1987learning} states that any regular language $L$ can be exactly learned with a number of queries 
that is polynomial in the number of states of a minimum deterministic automaton representing $L$, and in the size of the largest counter-example
returned by the teacher. Since an OBDD of length $n$ and width $w$ may be regarded as an acyclic finite automaton with $O(w\cdot n)$ states, and since one can assume in this case that counter-examples are of size $O(n)$, Angluin's theorem 
implies the following corollary. 

\begin{corollary}
\label{corollary:Angluin}
Let $L\subseteq \{0,1\}^n$ be a language with the property that there is an OBDD 
$\D\in \detCalB(\width)^{\circledast}$ with $\languageL = \languageL(\D)$. Then one can learn 
$L$ with $\width^{O(1)}\cdot n^{O(1)}$ membership and equivalence queries. 
\end{corollary}

Therefore, in view of Corollary \ref{corollary:Angluin} our results (Theorem \ref{theorem:MainTechnicalDecomposition}, Theorem \ref{theorem:UpperBoundsGenFunction} and Theorem \ref{corollary:Juntas}) can also be used to provide a parameterized approach for approaching the problems \pnFunctionDecomposition, \pnRecFunctionDecomposition and \pnGenJunta even if an explicit description of the target OBDD to be decomposed is not known, as long as we do have an oracle that is able to answer membership and equivalence queries. 

%% file: journalbib.bib
@article{shannon1949synthesis,
  author    = {Claude E. Shannon},
  title     = {The Synthesis of Two-Terminal Switching Circuits},
  journal   = {Bell System Technical Journal},
  volume    = {28},
  number    = {1},
  pages     = {59--98},
  year      = {1949},
  doi       = {10.1002/j.1538-7305.1949.tb03624.x}
}

@inproceedings{molnar2020quantifying,
  title={Quantifying model complexity via functional decomposition for better post-hoc interpretability},
  author={Molnar, Christoph and Casalicchio, Giuseppe and Bischl, Bernd},
  booktitle={Machine Learning and Knowledge Discovery in Databases: International Workshops of ECML PKDD 2019, W{\"u}rzburg, Germany, September 16--20, 2019, Proceedings, Part I},
  pages={193--204},
  year={2020},
  organization={Springer}
}

@article{bohanec2004function,
  title={A function-decomposition method for development of hierarchical multi-attribute decision models},
  author={Bohanec, Marko and Zupan, Bla{\v{z}}},
  journal={Decision Support Systems},
  volume={36},
  number={3},
  pages={215--233},
  year={2004},
  publisher={Elsevier}
}

@article{hashemi2018evolutionary,
  title={Evolutionary self-expressive models for subspace clustering},
  author={Hashemi, Abolfazl and Vikalo, Haris},
  journal={IEEE Journal of Selected Topics in Signal Processing},
  volume={12},
  number={6},
  pages={1534--1546},
  year={2018},
  publisher={IEEE}
}

@inproceedings{hiabu2023unifying,
  title={Unifying local and global model explanations by functional decomposition of low dimensional structures},
  author={Hiabu, Munir and Meyer, Joseph T and Wright, Marvin N},
  booktitle={International Conference on Artificial Intelligence and Statistics},
  pages={7040--7060},
  year={2023},
  organization={PMLR}
}

@inproceedings{laberge2024tackling,
  title={Tackling the XAI Disagreement Problem with Regional Explanations},
  author={Laberge, Gabriel and Pequignot, Yann Batiste and Marchand, Mario and Khomh, Foutse},
  booktitle={International Conference on Artificial Intelligence and Statistics},
  pages={2017--2025},
  year={2024},
  organization={PMLR}
}

@incollection{zupan1999function,
  title={Function decomposition in machine learning},
  author={Zupan, Bla{\v{z}} and Bratko, Ivan and Bohanec, Marko and Dem{\v{s}}ar, Janez},
  booktitle={Advanced Course on Artificial Intelligence},
  pages={71--101},
  year={1999},
  publisher={Springer}
}

@inproceedings{zupan1997machine,
  title={Machine learning by function decomposition},
  author={Zupan, Blaz and Bohanec, Marko and Bratko, Ivan and Demsar, Janez},
  booktitle={ICML},
  pages={421--429},
  year={1997}
}

@article{deniziak2021decolib,
  title={DECOLib: A library of components for DECOmposition of discrete functions},
  author={Deniziak, Stanis{\l}aw and Wi{\'s}niewski, Mariusz},
  journal={SoftwareX},
  volume={16},
  pages={100799},
  year={2021},
  publisher={Elsevier}
}

@article{vykhovanets2006algebraic,
  title={Algebraic decomposition of discrete functions},
  author={Vykhovanets, Valeriy Sviatoslavovich},
  journal={Automation and Remote Control},
  volume={67},
  number={3},
  pages={361--392},
  year={2006},
  publisher={Springer}
}

@article{bryant1992symbolic,
  title={Symbolic boolean manipulation with ordered binary-decision diagrams},
  author={Bryant, Randal E},
  journal={ACM Computing Surveys (CSUR)},
  volume={24},
  number={3},
  pages={293--318},
  year={1992},
  publisher={ACM New York, NY, USA}
}

@InProceedings{Kupferman2013,
author="Kupferman, Orna
and Mosheiff, Jonathan",
editor="Chatterjee, Krishnendu
and Sgall, Jir{\'i}",
title="Prime Languages",
booktitle="Mathematical Foundations of Computer Science 2013",
year="2013",
publisher="Springer Berlin Heidelberg",
address="Berlin, Heidelberg",
pages="607--618",
isbn="978-3-642-40313-2"
}

@article{Woelfel2006symbolic,
  title={Symbolic topological sorting with OBDDs},
  author={Woelfel, Philipp},
  journal={Journal of Discrete Algorithms},
  volume={4},
  number={1},
  pages={51--71},
  year={2006},
  publisher={Elsevier}
}

@inproceedings{Bollig2014width,
  title={On the width of ordered binary decision diagrams},
  author={Bollig, Beate},
  booktitle={International Conference on Combinatorial Optimization and Applications},
  pages={444--458},
  year={2014},
  organization={Springer}
}

@inproceedings{Hachtel1993symbolic,
  title={A symbolic algorithm for maximum flow in 0-1 networks},
  author={Hachtel, Gary D and Somenzi, Fabio},
  booktitle={Proceedings of the 1993 IEEE/ACM international conference on Computer-aided design},
  pages={403--406},
  year={1993},
  organization={IEEE Computer Society Press}
}

@inproceedings{Kuske21,
  author    = {Dietrich Kuske},
  title     = {Second-Order Finite Automata: Expressive Power and Simple Proofs Using
               Automatic Structures},
  booktitle = {Proc. of the 25th International Conference on Developments in Language Theory ({DLT} 2021)},
  pages     = {242--254},
  year      = {2021},
  editor    = {Nelma Moreira and
               Rog{\'{e}}rio Reis},
  series    = {Lecture Notes in Computer Science},
  volume    = {12811},
  publisher = {Springer},
}

@inproceedings{blumensath2000automatic,
  author    = {Achim Blumensath and
               Erich Gr{\"{a}}del},
  title     = {Automatic Structures},
  booktitle = {Proc. of the 15th Annual {IEEE} Symposium on Logic in Computer Science (LICS 2000)},
  pages     = {51--62},
  year      = {2000},
  publisher = {{IEEE} Computer Society},
}

@inproceedings{MeloOliveira2022,
	  author    = {Alexsander Andrade de Melo and
		                 Mateus de Oliveira Oliveira},
	    title     = {Second-Order Finite Automata},
	      booktitle = {Proc. of the  15th International Computer Science Symposium in Russia ({CSR} 2020)},
	        pages     = {46--63},
		  year      = {2020},
	  editor    = {Henning Fernau},
	      series    = {Lecture Notes in Computer Science},
	        volume    = {12159},
}

@inproceedings{MeloOliveira2019,
  author    = {Alexsander Andrade de Melo and
               Mateus de Oliveira Oliveira},
  title     = {On the Width of Regular Classes of Finite Structures},
  booktitle = {Automated Deduction - {CADE} 27 - 27th International Conference on
               Automated Deduction, Natal, Brazil, August 27-30, 2019, Proceedings},
  pages     = {18--34},
  year      = {2019},
  series    = {Lecture Notes in Computer Science},
  volume    = {11716},
  publisher = {Springer},
  year      = {2019},
}

@inproceedings{abu2017advice,
author    = {Faried Abu Zaid and Erich Gr{\"{a}}del and Frederic Reinhardt},
title     = {Advice Automatic Structures and Uniformly Automatic Classes},
pages     = {35:1--35:20},
year      = {2017},
editor    = {Valentin Goranko and Mads Dam},
booktitle = {Proc. of the 26th {EACSL} Annual Conference on Computer Science Logic ({CSL} 2017)},
series    = {LIPIcs},
volume    = {82},
publisher = {Schloss Dagstuhl - Leibniz-Zentrum f{\"{u}}r Informatik},
}

@article{jain2012learnability,
  title={Learnability of automatic classes},
  author={Jain, Sanjay and Luo, Qinglong and Stephan, Frank},
  journal={Journal of Computer and System Sciences},
  volume={78},
  number={6},
  pages={1910--1927},
  year={2012},
  publisher={Elsevier}
}

@incollection{hopcroft1971n,
    title={An n log n algorithm for minimizing states in a finite automaton},
    author={Hopcroft, John},
    booktitle={Theory of machines and computations},
    pages={189--196},
    year={1971},
    publisher={Elsevier}
}

@article{Bollig2012symbolic,
  title={On symbolic OBDD-based algorithms for the minimum spanning tree problem},
  author={Bollig, Beate},
  journal={Theoretical Computer Science},
  volume={447},
  pages={2--12},
  year={2012},
  publisher={Elsevier}
}

@inproceedings{Sawitzki2004implicit,
  title={Implicit flow maximization by iterative squaring},
  author={Sawitzki, Daniel},
  booktitle={International Conference on Current Trends in Theory and Practice of Computer Science},
  pages={301--313},
  year={2004},
  organization={Springer}
}

@inproceedings{jecker2020unary,
  title={Unary prime languages},
  author={Jecker, Ismael R and Kupferman, Orna and Mazzocchi, Nicolas},
  booktitle={45th International Symposium on Mathematical Foundations of Computer Science},
  volume={170},
  year={2020}
}

@inproceedings{de2021robust,
  title={Robust testing of low dimensional functions},
  author={De, Anindya and Mossel, Elchanan and Neeman, Joe},
  booktitle={Proceedings of the 53rd Annual ACM SIGACT Symposium on Theory of Computing},
  pages={584--597},
  year={2021}
}

@inproceedings{de2019your,
  title={Is your function low dimensional?},
  author={De, Anindya and Mossel, Elchanan and Neeman, Joe},
  booktitle={Conference on Learning Theory},
  pages={979--993},
  year={2019},
  organization={PMLR}
}

@article{blais2010testing,
  title={Testing juntas: A brief survey},
  author={Blais, Eric},
  journal={Property Testing: Current research and surveys},
  pages={32--40},
  year={2010},
  publisher={Springer}
}

@inproceedings{DBLP:conf/concur/JeckerM021,
  author       = {Isma{\"{e}}l Jecker and
                  Nicolas Mazzocchi and
                  Petra Wolf},
  editor       = {Serge Haddad and
                  Daniele Varacca},
  title        = {Decomposing Permutation Automata},
  booktitle    = {32nd International Conference on Concurrency Theory, {CONCUR} 2021,
                  August 24-27, 2021, Virtual Conference},
  series       = {LIPIcs},
  volume       = {203},
  pages        = {18:1--18:19},
  publisher    = {Schloss Dagstuhl - Leibniz-Zentrum f{\"{u}}r Informatik},
  year         = {2021},
  url          = {https://doi.org/10.4230/LIPIcs.CONCUR.2021.18},
  doi          = {10.4230/LIPIcs.CONCUR.2021.18},
  bibsource    = {dblp computer science bibliography, https://dblp.org}
}

@preamble{
   "\def\cprime{$'$} "
}

@article{Case2013,
  title   = {Automatic functions, linear time and learning},
  author  = {John Case and Sanjay Jain and Frank Stephan},
  journal = {Logical Methods in Computer Science},
  colume  = {Volume 9, Issue 3},
  year    = {2013},
}

@article{DBLP:journals/jair/Srivastava23,
  author       = {Siddharth Srivastava},
  title        = {Hierarchical Decompositions and Termination Analysis for Generalized
                  Planning},
  journal      = {J. Artif. Intell. Res.},
  volume       = {77},
  pages        = {1203--1236},
  year         = {2023},
  url          = {https://doi.org/10.1613/jair.1.14185},
  doi          = {10.1613/JAIR.1.14185},
  bibsource    = {dblp computer science bibliography, https://dblp.org}
}

@inproceedings{DBLP:conf/ijcai/GoelC89,
  author       = {Ashok K. Goel and
                  B. Chandrasekaran},
  editor       = {N. S. Sridharan},
  title        = {Functional Representation of Designs and Redesign Problem Solving},
  booktitle    = {Proceedings of the 11th International Joint Conference on Artificial
                  Intelligence. Detroit, MI, USA, August 1989},
  pages        = {1388--1394},
  publisher    = {Morgan Kaufmann},
  year         = {1989},
  url          = {http://ijcai.org/Proceedings/89-2/Papers/086.pdf},
  bibsource    = {dblp computer science bibliography, https://dblp.org}
}

@inproceedings{DBLP:conf/ijcai/SunNM19a,
  author       = {Lu Sun and
                  Canh Hao Nguyen and
                  Hiroshi Mamitsuka},
  editor       = {Sarit Kraus},
  title        = {Multiplicative Sparse Feature Decomposition for Efficient Multi-View
                  Multi-Task Learning},
  booktitle    = {Proceedings of the Twenty-Eighth International Joint Conference on
                  Artificial Intelligence, {IJCAI} 2019, Macao, China, August 10-16,
                  2019},
  pages        = {3506--3512},
  publisher    = {ijcai.org},
  year         = {2019},
  url          = {https://doi.org/10.24963/ijcai.2019/486},
  doi          = {10.24963/IJCAI.2019/486},
  bibsource    = {dblp computer science bibliography, https://dblp.org}
}

@inproceedings{DBLP:conf/ijcai/IzzaJRB20,
  author       = {Yacine Izza and
                  Sa{\"{\i}}d Jabbour and
                  Badran Raddaoui and
                  Abdelhamid Boudane},
  editor       = {Christian Bessiere},
  title        = {On the Enumeration of Association Rules: {A} Decomposition-based Approach},
  booktitle    = {Proceedings of the Twenty-Ninth International Joint Conference on
                  Artificial Intelligence, {IJCAI} 2020},
  pages        = {1265--1271},
  publisher    = {ijcai.org},
  year         = {2020},
  url          = {https://doi.org/10.24963/ijcai.2020/176},
  doi          = {10.24963/IJCAI.2020/176},
  bibsource    = {dblp computer science bibliography, https://dblp.org}
}

@article{correa2023humans,
  title={Humans decompose tasks by trading off utility and computational cost},
  author={Correa, Carlos G and Ho, Mark K and Callaway, Frederick and Daw, Nathaniel D and Griffiths, Thomas L},
  journal={PLoS computational biology},
  volume={19},
  number={6},
  pages={e1011087},
  year={2023},
  publisher={Public Library of Science San Francisco, CA USA}
}

@inproceedings{DBLP:conf/nips/AtzmonKSC20,
  author       = {Yuval Atzmon and
                  Felix Kreuk and
                  Uri Shalit and
                  Gal Chechik},
  editor       = {Hugo Larochelle and
                  Marc'Aurelio Ranzato and
                  Raia Hadsell and
                  Maria{-}Florina Balcan and
                  Hsuan{-}Tien Lin},
  title        = {A causal view of compositional zero-shot recognition},
  booktitle    = {Advances in Neural Information Processing Systems 33: Annual Conference
                  on Neural Information Processing Systems 2020, NeurIPS 2020, December
                  6-12, 2020, virtual},
  year         = {2020},
  url          = {https://proceedings.neurips.cc/paper/2020/hash/1010cedf85f6a7e24b087e63235dc12e-Abstract.html},
  bibsource    = {dblp computer science bibliography, https://dblp.org}
}

@inproceedings{DBLP:conf/ijcai/WangYWD23,
  author       = {Henan Wang and
                  Muli Yang and
                  Kun Wei and
                  Cheng Deng},
  title        = {Hierarchical Prompt Learning for Compositional Zero-Shot Recognition},
  booktitle    = {Proceedings of the Thirty-Second International Joint Conference on
                  Artificial Intelligence, {IJCAI} 2023, 19th-25th August 2023, Macao,
                  SAR, China},
  pages        = {1470--1478},
  publisher    = {ijcai.org},
  year         = {2023},
  url          = {https://doi.org/10.24963/ijcai.2023/163},
  doi          = {10.24963/IJCAI.2023/163},
  bibsource    = {dblp computer science bibliography, https://dblp.org}
}

@article{angluin1987learning,
	  title={Learning regular sets from queries and counterexamples},
	    author={Angluin, Dana},
	      journal={Information and computation},
	        volume={75},
		  number={2},
		    pages={87--106},
		      year={1987},
		        publisher={Elsevier}
}
